\documentclass[12pt]{article}

\usepackage{amsmath}
\usepackage{amssymb}
\usepackage{theorem}
\theorembodyfont{\rmfamily}
\usepackage[mathscr]{eucal}
\usepackage{comment}
\newtheorem{theorem}{Theorem}[section]
\newtheorem{df}[theorem]{\bf Definition}
\newtheorem{thm}[theorem]{\bf Theorem}

\newtheorem{lem}[theorem]{\bf Lemma}

\newtheorem{prop}[theorem]{\bf Proposition}
\newtheorem{assumption}[theorem]{\bf Assumption}

\newtheorem{pf}{\it Proof:}

\newcommand{\hf}{H_{\rm f}}
\newcommand{\hfm}{H_{{\rm f},m}}
\newcommand{\hhfm}{\hat H_{{\rm f},m}}
\newcommand{\eq}[1]{\begin{equation}\label{#1}}
\newcommand{\en}{\end{equation}}
\newcommand{\kak}[1]{(\ref{#1})}

\newcommand{\dis}{\displaystyle}
\newcommand{\qed}
{\hfill\hbox{\rule{6pt}{6pt}}} 

\makeatletter
\@addtoreset{equation}{section}
\makeatother

\title{\sc Existence of the ground state for the Pauli-Fierz model with a variable mass}
\author{Takeru Hidaka\\
Faculty of Mathematics, Kyushu University,\\
Fukuoka, Japan, 819-0395}
\date{}
\begin{document}
 \setlength{\baselineskip}{16pt}
\maketitle
\begin{abstract}
The Pauli-Fierz model with a variable mass $v$ is considered.
An ultraviolet cutoff and an infrared regularity condition are imposed on
a quantized radiation field.
It is
shown that
the ground state exists
for arbitrary values of coupling
constants, when
$v(x)\leq c\langle x\rangle^{-\beta}$, $\beta>3$.
\end{abstract}
\section{Introduction}
A  Hamiltonian in quantum physics
describes the energy of a system and is realized as a self-adjoint operator on a Hilbert space.
The bottom of the spectrum of the Hamiltonian is called the ground state energy.
If the ground state energy is an eigenvalue, the eigenvector
associated with this is  called a ground state.

The Pauli-Fierz Hamiltonian describes low energy electrons minimally
coupled to a quantized radiation field.
In this paper we consider a single spinless electron coupled to a quantized radiation field with a variable mass.
Throughout an ultraviolet cutoff is imposed on the quantized radiation field and an infrared regularity condition is also done.
We show the existence of ground states for arbitrary values of coupling constants when the variable mass decays sufficiently fast.
This paper is inspired by \cite{GHPS,GHPS2,GHPS3}, where some scalar quantum field model on a pseudo Riemannian  manifold is considered and properties of the ground state are discussed.

{\bf (The Pauli-Fierz Hamiltonian)}
The standard Pauli-Fierz Hamiltonian $H_{\text{PF}}$ can be realized as a self-adjoint operator on
the Hilbert space $\mathcal{H}_{P}\otimes \mathcal{F}$, where
$\mathcal{H}_{P}=L^{2}(\mathbb{R}^{3})$ and
$\mathcal{F}$ is the boson Fock space over $L^{2}(\mathbb{R}^{3};\mathbb{C}^{2})$.
$\mathcal{H}_{P}$ denotes the state space for a single spinless electron and $\mathcal{F}$ for photons quantized in the Coulomb gauge.
Formally, $H_{\text{PF}}$ is described as
\eq{5}
H_{\text{PF}}=\frac{1}{2}
\left(p+\sqrt{\alpha}A\right)^{2}+\hf +V,
\en
where $p=(p_{1},p_{2},p_{3})=(-i\partial_1,-i\partial_2,-i\partial_3)$ denotes the momentum operator of an electron, $0\leq\alpha$ a coupling constant,  $V$ an external potential,
$A=(A_{1},A_{2},A_{3})$
a quantized radiation field with an ultraviolet cutoff, and $\hf $ the free field Hamiltonian on $\mathcal {F}$.
The creation operator and the annihilation operator in $\mathcal F$ smeared by
$f\in L^{2}(\mathbb{R}^{3})$ are  denoted by
$a^{\dagger}(f)$ and $a(f)$,
respectively.
The  $\mu$th component of
the quantized radiation
field
$A_{\mu}(x)$ is given by the sum of a creation operator and an annihilation operator:
\eq{1}
A_{\mu}(x)=\frac{1}{\sqrt{2}}\left(a^{\dagger}\left(\frac{ \hat{\varphi}e_{\mu}e^{-ikx} }{ \sqrt{\omega} }\right)+a\left(\frac{ \overline{\hat{\varphi}}e_{\mu}e^{ikx} }{ \sqrt{\omega} }\right)\right)
\en
for each $x\in\mathbb{R}^{3}$.
 Here $\hat{\varphi}$ is a cutoff function, $e_\mu$  the $\mu$th component of polarization vectors and  \eq{30}
 \omega(k)=|k|
 \en
 describes the energy of one  photon with momentum $k\in\mathbb R^3$ and is called a dispersion relation.
Finally the free field Hamiltonian
$\hf $ on $\mathcal F$ is given by the second quantization of $\omega$:
\eq{7}
\hf =d\Gamma(\omega).
\en

It is shown that $H_{\text{PF}}$ is self-adjoint by the Kato-Rellich theorem
if the coupling constant $\alpha$ is sufficiently small.
The self-adjointness of $H_{PF}$ for an arbitrary values of the coupling constant is shown in \cite{HH,hi00,hi02}.
In our paper, the Hamiltonian is however defined by the unique self-adjoint operator associated with a semi-bounded quadratic form.

The existence of the ground state for the Pauli-Fierz Hamiltonian (or related models) is investigated in \cite{A,BFS,G,GLL,hi99,HS,hi01,Sp98}.
When $V(x)=|x|^{2}$, the existence and uniqueness of the ground state is proven by Arai \cite{A} for  sufficiently small values of a coupling constant in the dipole-approximation.
By Hiroshima \cite{hi99,hi01}, for sufficiently small values of coupling constants, the existence of the ground state of $H_{PF}$ is proven without the dipole-approximation but the infrared regularity condition is supposed.
Without the infrared regularity condition, the existence of the ground state of $H_{\text{PF}}$ is shown by Bach, Fr\"ohlich and Sigal \cite{BFS}.
G\'erard and Spohn \cite{G,Sp98} show that for arbitrary values of a coupling constant, the ground state exists for a related model.
Without the infrared regularity condition and for all values of coupling constant, the existence of the ground state of $H_{\text{PF}}$ is proven by Griesemer, Lieb and Loss \cite{GLL}.
The uniqueness of the ground state of $H_{\text{PF}}$ is shown in \cite{hi9}.
We extend this to a version with a variable mass.

{\bf (The Pauli-Fierz Hamiltonian with a variable mass)}
From now on we are concerned with  the Pauli-Fierz Hamiltonian but with a variable mass.
 As is seen in the standard Pauli-Fierz Hamiltonian \kak{5} the test function of
 $A_\mu(x)$ is given by
\eq{20}
e_\mu(k)e^{-ikx}\hat\varphi(k)/\sqrt{\omega(k)}
\en
 in the momentum representation. See \kak{1}.
Since the position representation of the dispersion relation $\omega$ is $\hat\omega
=\sqrt{-\Delta}$,
the test function \kak{20}
is
given by
\eq{8}
\rho_x(y)=({-\Delta_y})^{-1/2}
\left(
(2\pi)^{-3/2}
\int e_\mu(k)e^{-ikx} e^{+ik y }\hat\varphi(k) dk\right)
\en
in the position representation,
and it follows that
\eq{9}
(-\Delta_x) e^{ikx}=|k|^2e^{ikx}.
\en
Then $H_{\rm PF}$ is expressed
in the position representation as
\eq{40}
\frac{1}{2}(p+\sqrt\alpha A(\rho_x))^2+d\Gamma(\sqrt{-\Delta})+V,\en
where
\eq{41}
A(\rho_x)=\frac{1}{\sqrt 2}(a^\dagger(\rho_x)+a(\overline{\rho_x})).
\en
The Pauli-Fierz Hamiltonian with a variable mass is defined by
$H_{\text {PF}}$ with $\hat \omega=\sqrt{-\Delta}$
and $e^{ikx}$ replaced by $\sqrt{-\Delta +v}$ with some real-valued function $v$
and $\Psi(k,x)$, respectively.
Here $\Psi(k,x)$ is a solution to the equation corresponding  to \kak{9}:
\eq{11}
(-\Delta_x+v(x)) \Psi(k,x)=|k|^2\Psi(k,x),
\en
and $v$ is called a variable mass.
Thus the Pauli-Fierz Hamiltonian with a variable mass
is defined by
\eq{42}
\frac{1}{2}(p+\sqrt\alpha A(\rho_x^\Psi))^2+
d\Gamma(\sqrt{-\Delta+v})+V\en
in the position representation.
Here
the test function of $A_\mu$ is
given by
\eq{12}
\rho_x^\Psi(y)=(-\Delta_y+v(y))^{-1/2}
\left(
{
(2\pi)^{-3/2}
\int e_\mu(k)
\overline{\Psi(k,x)}{\Psi(k,y)}
\hat \varphi(k) dy}\right).
\en
A sufficient condition such that
a solution $\Psi(k,x)$ to equation \kak{11} exists
is known in \cite{Ik}.
Suppose that \begin{eqnarray}
\label{2}
&&v(x)=\mathcal{O}(|x|^{-\beta}),\quad \beta >3,\\
&&
\label{3}
\sigma_{P}(-\Delta+v)\subset (0,\infty),
\end{eqnarray}
where $\sigma_{P}(-\Delta+v)$
denotes the set of eigenvalues of $-\Delta+v$.
Then the generalized eigenfunction $\Psi(k,x)$ satisfying \kak{11}
exists and
it obeys
the so-called Lippman-Schwinger equation:
\eq{15}
\Psi(k,x)=e^{ikx}-\frac{1}{4\pi}\int\frac{e^{i|k||x-y|}v(y)}{|x-y|}\Psi(k,y)dy.
\en

{\bf (Unitary transformation)}
The Pauli-Fierz Hamiltonian with a variable mass \kak{42}
can be transformed to a convenient form under some condition on $v$. Under \kak{2} and \kak{3}, the generalized Fourier transformation $\mathscr{F} :L^2(\mathbb R^3)\to L^2(\mathbb R^3)$ can be defined
through $\Psi(k,x)$ by
\eq{44}
\mathscr{F} f(k)=(2\pi )^{-3/2}\mathrm{l.i.m.}\int f(x) \overline{\Psi (k,x)} dx. \en
Since
\eq{21}
\mathscr{F}(-\Delta+v)^{1/2}
\mathscr{F}^{-1}=|k|
\en
and
\eq{21-0}
(\mathscr{F}\rho_x^\Psi)(k)=e_\mu(k)\overline{\Psi(k,x)}
\hat\varphi(k)/\sqrt{\omega(k)},
\en
the Pauli-Fierz Hamiltonian with
a variable mass \kak{42}
can be
unitarily transformed
to the standard one
but with $e^{ikx}$ replaced by $\overline{\Psi(k,x)}$.
Thus our Hamiltonian is finally of the form
\eq{6}
H^V=\frac{1}{2}
\left(p+\sqrt{\alpha}A^\Psi\right)^{2}+\hf +V,
\en
where
\eq{28}
A^\Psi_{\mu}(x)=\frac{1}{\sqrt{2}}\left(a^{\dagger}\left(\frac{ \hat{\varphi}e_{\mu}\overline{
\Psi(\cdot,x)}}{ \sqrt{\omega} }\right)+a\left(\frac{ \overline{\hat{\varphi}}e_{\mu}
\Psi(\cdot,x)}{ \sqrt{\omega} }\right)\right)
\en
 in the momentum representation.
In this paper more general test functions than that of \kak{28} are treated. See \kak{33}.

{\bf (Outline of strategy)}
We show an outline of the strategy to  prove  the existence of ground states.

Let ${\omega}_{m}=\sqrt{|k|^2+m^{2}}$ and $H_{m}^{V}$ be defined by $H^{V}$ with $\hf $ replaced by $\hfm =d\Gamma(\hat{\omega}_{m})$.
The key facts are the Lippman-Schwinger equation \kak{15} and Lemma~\ref{Psi}, where
some regularity
 properties of $\Psi(k,x)$ is shown and  difference between $e^{ikx}$ and $\Psi(k,x)$ are estimated.

First we show the existence of ground states of $H_{m}^{V}$ by a  localization estimate.
In this estimate, we need the continuity of $\Psi(k,x)$ with respect to $x$
and the bound
\eq{16}
|\Psi(k,x)-e^{ikx}|\leq \frac{C}{\sqrt{1+|x|^2}}.
\en
This bound is proven in
  (a) of Lemma \ref{Psi}.

Secondly we show that as $m$ goes to zero, a normalized ground state of $H_{m}^{V}$ {\it strongly} converges to some non-zero vector, and this is a ground state of $H^V$.
In order to show this we need
to show a spatial exponential decay of a ground state of $H_m^V$ {\it uniformly} in $m\geq0$ and
two bounds:
a photon number bound
(Lemmas \ref{lem of p.t.} and \ref{number b'dd})
 and
a photon derivative bound
 (Lemma \ref{derivative b'dd}).
To derive the former bound,
we show that $\Psi(k,x)$ is differentiable with respect to $x$ for each $k$ but $k\not=0$ and
 \eq{31}
 \sup_{k\in D ,x\in \mathbb{R}^{3}} \left| \frac{\partial \Psi}{\partial x_{\mu}}(k,x)\right|<\infty\en
  for any compact set $D\not \ni0$.
 This is derived from
\eq{43}
{\frac{\partial \Psi}{\partial x_{\mu}}(k,x)}
=ik_{\mu}e^{ikx}-
\frac{1}{4\pi}\int_{\mathbb{R}^{3}} \left(\frac{1}{|x-y|}
-{i|k|}\right)
\frac{(x_{\mu}-y_{\mu})e^{i|k||x-y|}
v(y)}{|x-y|^2}
\Psi(k,y)dy,
\en
which is shown in (b) of Lemma \ref{Psi},
while a photon derivative bound is derived from the bounds
\begin{eqnarray}
\frac{1}{|h|}|\Psi(k+h,x)-\Psi(k,x)|
&\leq& C(1+|x|),\\
\frac{1}{|h|}\left| \frac{\partial \Psi}{\partial x_{\nu}}(k+h,x)-\frac{\partial \Psi}{\partial x_{\nu}}(k,x)\right|&\leq& C(1+|k|+|x|+|k| |x|).
\end{eqnarray}
These bounds are also shown in (c) of Lemma \ref{Psi}.

Finally a spatial exponential decay of a ground state of $H_m^V$ is proven in \cite{FH} by means of functional integral representations, by which,
furthermore, the uniqueness of the ground state
is also shown.

{\bf (Infrared regularity  condition)}
We give a comment on the infrared regularity condition for the Pauli-Fierz model with a variable mass.
The existence of a ground state for the standard Pauli-Fierz Hamiltonian is proven without the infrared regularity condition in \cite{BFS,GLL}.
One of the key fact is
the bound
$
|e^{ikx}-1|\leq |k||x|$.
In our case however
\eq{18}
|\Psi(k,x)-1|\leq |k||x|\en
does {\it not} hold generally.
Then
unfortunately
infrared regularity condition
\eq{19}
\int_{\mathbb{R}^{3}}\frac{|\hat{\varphi}(k)|^{2p}}
{\omega(k)^{5p}}dk<\infty,\quad 0\leq \forall p<1,
\en
is supposed.

{\bf (Comparing with a scalar model)}
The ground state of some scalar quantum field model, which is called the Nelson model,  with a variable mass is studied in \cite{GHPS}, where
a variable mass is derived from a wave equation on a pseudo Riemannian manifold.
The Nelson Hamiltonian on a pseudo Riemannian manifold is given by
\eq{22}
H_{\text{Nelson}}=H_{P}\otimes 1 + 1\otimes \hf+ \alpha\Phi,
\en
where $H_{P}$ is a Schr\"odinger operator
  and interaction $\Phi$
is given by
\eq{23}
\Phi(x)=\frac{1}{\sqrt{2}}\left(a^{\dagger}\left(\hat{\omega}^{-1/2}\rho_{x}\right)+a\left(\overline{\hat{\omega}^{-1/2}\rho_{x}}\right)\right)
\en
for each $x$
in the position representation.
Here a dispersion relation $\hat{\omega}$ is given by
$\hat{\omega}=\sqrt{-\Delta+v}$
with a variable mass $v$ and
\eq{24}
\rho_{x}(y)=(2\pi)^{-3/2}\int \Psi(k,y)\overline{\Psi(k,x)}\hat{\varphi}(k)dk
\en
with
a cutoff function
$\hat{\varphi}(k)$
and  the generalized eigenfunction
$\Psi(k,x)$ satisfying
\eq{25}
(-\Delta+v)\Psi(k,x)=|k|^{2}\Psi(k,x).
\en
When a variable mass $v$ is
$v(x)=\mathcal{O}(|x|^{-\beta})$ with $\beta >3$ and
$
\hat\varphi(0)>0$,
it is shown in \cite{GHPS} that $H_{\text{Nelson}}$ has no ground states. Note that
$\hat\varphi(0)>0$
implicitly implies that
\eq{29}
\int\frac{|\hat\varphi(k)|^2}
{\omega(k)^3} dk=\infty.
\en
In this paper however we assume \kak{19} and hence the integral in \kak{29} is concequently finite.

{\bf (Organization)}
This paper is organized as follows:
Section 2 is devoted to defining the Pauli-Fierz Hamiltonian by a quadratic form and showing regularity properties of $\Psi(k,x)$.
In Section 3 we mention the main theorem.
In Section 4 we show the existence of a ground state for the massive Hamiltonian, and
in Section 5,
for the massless Hamiltonian.
Finally Section 6 is devoted to giving appendices.

\section{Definition}
\subsection{Definition of the Pauli-Fierz model}
The Pauli-Fierz Hamiltonian is defined as a self-adjoint operator on the Hilbert space:
\begin{eqnarray}
\mathcal{H}=\mathcal{H}_{p}\otimes \mathcal{F},
\end{eqnarray}
where $\mathcal{H}_{p}=L^{2}(\mathbb{R}^{3})$ describes the state space for a single electron, and $\mathcal{F}$ the boson Fock space over $L^{2}(\mathbb{R}^{3};\mathbb{C}^{2})$, which is defined by
\begin{eqnarray}
\mathcal{F}&=
&\bigoplus_{n=0}^{\infty}\otimes_s^{n}\,L^{2}(\mathbb{R}^{3};\mathbb{C}^{2}) \nonumber \\
&=&\left\{\{\Psi^{(n)}\}_{n=0}^{\infty}\Big|\Psi^{(n)}\in \otimes_{s}^{n}L^{2}(\mathbb{R}^{3};\mathbb{C}^{2}),\;\sum_{n=0}^{\infty}\Vert\Psi^{(n)}\Vert^{2}<\infty \right\}.
\end{eqnarray}
Here $\dis \otimes_s^{n}\,L^{2}(\mathbb{R}^{3};\mathbb{C}^{2})$ denotes the $n$-fold symmetric tensor product of $L^{2}(\mathbb{R}^{3};\mathbb{C}^{2})$ with $\dis \otimes_{s}^{0}L^{2}(\mathbb{R}^{3};\mathbb{C}^{2})=\mathbb{C}$.
$\mathcal{F}$ describes the state space for photons. The inner product on $\mathcal{F}$ is defined by
\begin{eqnarray*}
(\,\Psi,\,\Phi\,)_{\mathcal{F}}&=&\sum_{n=0}^{\infty} (\,\Psi^{(n)},\,\Phi^{(n)}\,)_{ \otimes_{s}^{n} L^{2} (\mathbb{R}^{3};\mathbb{C}^{2}) }\\
&=&\overline{\Psi^{(0)}}\Phi^{(0)}+\sum_{n=1}^{\infty}\int \overline{\Psi^{(n)}(k_{1},\ldots,k_{n})}\Phi^{(n)}(k_{1},\ldots, k_{n})dk_{1}\cdots dk_{n}.
\end{eqnarray*}
The finite particle subspace of $\mathcal{F}$ is given by
$$
\mathcal{F}_{0}=\{\{\Psi^{(n)}\}_{n=0}^{\infty}\mathcal{F}\,|\,\Psi^{(n)}=0, n\geq n_{0}\text{ with some }n_{0}\, \}.
$$
In this paper, $\mathcal{H}$ can be identified with the set of $\mathcal{F}$-valued $L^{2}$-functions on $\mathbb{R}^{3}$:
\[ \mathcal{H} \simeq 
\int^{\oplus}_{\mathbb{R}^{3}}\mathcal{F}\,dx.\]
Under this identification, the inner product on $\mathcal{H}$ is given by
\begin{eqnarray}
(\Psi,\Phi)_{\mathcal{H}} = \int_{\mathbb{R}^{3}} ( \Psi(x),\Phi(x))_{\mathcal{F}} dx.
\end{eqnarray}
Alternatively, $\mathcal{H}$ can be also identified with
\begin{eqnarray}
\lefteqn{\Big\{\Psi=\{\Psi^{(n)}\}_{n=0}^{\infty}\in L^{2}(\mathbb{R}^{3}) \oplus\bigoplus_{n=1}^{\infty}L^{2}(\mathbb{R}_x^3\times \mathbb{R}_k^{3n};\mathbb{C}^{2})}\nonumber \\
&& \Big|
\Psi^{(n)}(x,k_{\sigma (1)},\ldots,k_{\sigma (n)})=\Psi^{(n)}(x,k_{1},\ldots,k_{n}),
\sigma \in \mathfrak{S}_{n}, n\geq 1 \Big\},
\end{eqnarray}
where $\mathfrak{S}_{n}$ denotes the set of permutations of degree $n$.
Let $T$ be a densely defined closable operator on $L^{2}(\mathbb{R}^{3};\mathbb{C}^{2})$.
Then the second quantization of $T$ is defined by
\begin{eqnarray}
\Gamma(T)&=&\oplus_{n=0}^{\infty} \otimes^{n} T,\\
d\Gamma(T)&=&\oplus_{n=0}^{\infty} \otimes^{n} T^{(n)},
\end{eqnarray}
where
$\dis\otimes^{0}T=1$, $T^{(n)}=\overline {\sum_{k=1}^{n} 1\otimes\cdots 1\otimes \stackrel{k th}{T}\otimes 1\cdots \otimes 1}$
and $T^{(0)}=0$ and $\overline{S}$ denotes the closure of $T$.
The number operator is defined by the second quantization of the identity:
\[N=d\Gamma(1).\]
The annihilation operator $a(f)$ and the creation operator $a^{\dagger}(f)$ smeared by \\
$f\in L^{2}(\mathbb{R}^{3};\mathbb{C}^{2})$ on $\mathcal{F}$ are defined by
\begin{eqnarray}
&&D\left( a^{\dagger}(f)\right) =\left\{ \Psi\in\mathcal{F}\;\Big| \; \sum_{n=1}^{\infty} n\left\Vert S_{n}(f\otimes \Psi^{(n-1)}) \right\Vert^{2}<\infty \,\right\}, \\
&&\left(a^{\dagger}(f)\Psi\right)^{(n)}=\sqrt{n}S_{n}(f\otimes \Psi^{(n-1)}),n\geq 1,\quad\left(a^{\dagger}(f)\Psi\right)^{(0)}=0,\label{creation}\\
&&a(f)=(a^{\dagger}(\overline{f}))^{*}, \label{annihilation}
\end{eqnarray}
where $S_{n}$ is the symmetrization operator of degree $n$ and $D(T)$ denotes the domain of $T$.
$\Omega=(1,0,0,\cdots)\in \mathcal{F}$ is called the Fock vacuum. By $(\ref{creation})$ and $(\ref{annihilation})$,
$a(f)\Omega =0$ holds.
On the finite particle subspace $\mathcal{F}_{0}$, the annihilation operator and the creation operator satisfy canonical commutation relations:
\begin{eqnarray}
[\,a(f),\,a(g)\,]=0,\,\;\; [\,a^{\dagger}(f),\,a^{\dagger}(g)\,]=0, \,\;\; [\,a(f), \,a^{\dagger}(g)\,]=(\,\overline{f},\,g\,).
\end{eqnarray}
Let $T$ be a densely defined closable operator and $f\in D(T)$.
Then on the set
\begin{eqnarray}
\mathcal{L}\left(\{\Omega,a^{\dagger}(f_{1}) \cdots a^{\dagger} (f_{n})\Omega | f_{j}\in D(T) ,\;n\in \mathbb{N},\;j=1,\cdots ,n \}  \right) \label{N}
\end{eqnarray}
the following commutation relation follows:
\begin{eqnarray}
[\,d\Gamma(T),\,a^{\dagger}(f)\,]=a^{\dagger}(Tf),
\end{eqnarray}
where $\mathcal{L}(\{\cdots\})$ denotes the linear hull of $\{\cdots \}$.
Moreover, if $f\in D(T^{*})$,
the following commutation relation also follows on (\ref{N}):
\begin{eqnarray}
[\,d\Gamma(T),\, a(f)\,]=-a(T^{*}f).
\end{eqnarray}
Let $T$ be a nonnegative self-adjoint operator and $\ker T =\{0\}$.
Then for $f\in D(T^{-1/2})$ and $\Psi\in D(d\Gamma (T)^{1/2})$, $\Psi \in D(a(f))\cap D(a^{\dagger}(f))$ and
\begin{eqnarray}
\Vert a(f)\Psi \Vert&\leq& \Vert T^{-1/2}\Psi \Vert \,\Vert d\Gamma(T)^{1/2}\Psi \Vert, \label{b'dd1}\\
\Vert a^{\dagger}(f)\Psi \Vert&\leq& \Vert T^{-1/2}\Psi \Vert \,\Vert d\Gamma(T)^{1/2}\Psi \Vert +\Vert f\Vert \,\Vert \Psi \Vert\label{b'dd2}
\end{eqnarray}
hold.
For $\Psi\in D(N^{1/2})$, put
\begin{eqnarray}
(a_{1}\Psi(k))^{(n)}&=&\left( \sqrt{n+1}\,\Psi_{1}^{(n-1)}(k,k_{1},\cdots,k_{n}),\,0\right),\label{N1}\\
(a_{2}\Psi(k))^{(n)}&=&\left( 0,\,\sqrt{n+1}\,\Psi_{2}^{(n-1)}(k,k_{1},\cdots,k_{n})\right), \label{N2}\\
(a\Psi(k))^{(n)}(k_{1},\cdots,k_{n})&=&(a_{1}\Psi(k))^{(n)}+(a_{2}\Psi(k))^{(n)}.\label{a(k)}
\end{eqnarray}
Then for $\mathrm{a.e.}\, k$, $a\Psi(k)=\left\{(a\Psi(k))^{(n)} \right\}_{n=0}^{\infty}$ is a vector in $\mathcal{F}$.
We introduce assumptions on variable mass $v$:
\begin{assumption}\label{v}
Suppose that $v$ is a real-valued function on $\mathbb{R}^{3}$ such that
\begin{enumerate}
\item  $v(x)\leq C \left<x\right>^{-\beta}$ with $\beta>3$ and some constant $C\in (0,2)$, where $\left< x \right>=\sqrt{1+|x|^{2}}$;
\item  $-\Delta+v$ has no non-positive eigenvalues.
\end{enumerate}
\end{assumption}
\begin{df}[Dispersion relation]
The dispersion relation with a variable mass $v(x)$ is defined by
\begin{eqnarray}
\hat{\omega}=\sqrt{-\Delta_{x}+v(x) }.
\end{eqnarray}
\end{df}
Under Assumption $\ref{v}$, there exists the unique function
$\Psi(k,x)$ on $\left(\mathbb{R}^{3}\setminus\{0\}\right)\times \mathbb{R}^{3}$ such that
\eq{13}
\left( -\Delta_{x}+v(x)\right) \Psi(k,x)= |k|^{2}\Psi(k,x)
\en
and $\Psi(k,x)$ satisfies the Lippman-Schwinger equation:
\eq{Lip-Sch}
\Psi(k,x)=e^{ikx}-\frac{1}{4\pi}\int \frac{e^{i|k||x-y|}v(y)}{|x-y|}\Psi(k,y) dy.
\en
Put
\eq{33}
\rho_{x}^{\mu ,j}(y)= (2\pi)^{-3/2}\int \overline{\Psi(k,x)}\Psi(k,y)\hat{\varphi}_{j}^{\mu}(k)dk,\quad j=1,2.
\en
Here $\hat{\varphi}_{j}^{\mu}$ is a cutoff function.
Let us introduce assumptions on $\hat{\varphi}_{j}^{\mu}$.
\begin{assumption}\label{varphi}
\begin{enumerate}
\item  $\hat{\varphi}_{j}^{\mu}(k)$ is differentiable almost everywhere $k$ and
\begin{eqnarray}
\int_{ \mathbb{R}^{3} }\frac{ |\partial_{\lambda} \hat{\varphi}_{j}^{\mu}(k) |^{2p} }{ |k|^{3p} }dk<\infty  \text{ for all }0<p<1,\;\lambda=1,2,3; \label{IF}
\end{eqnarray}
\item {\bf (Infrared regularity condition)}
\begin{eqnarray}
\int_{\mathbb{R}^{3}}\frac{|\hat{\varphi}_{j}^{\mu}(k)|^{2p}}{|k|^{5p}}dk<\infty \text{ for all }0<p<1. \label{I.F.}
\end{eqnarray}
\item
\begin{eqnarray}
\int_{ \mathbb{R}^{3} } |k||\hat{\varphi}_j^\mu(k)|^2 dk<\infty.
\end{eqnarray}
\end{enumerate}
\end{assumption}
The quantized radiation field with a variable mass is defined as follows:
\begin{df}[Quantized radiation field]
For each $x\in \mathbb{R}^{3}$, we define
\begin{eqnarray}
A_{\mu}(x)=\frac{1}{\sqrt{2}}\left( a^{\dagger}\left(  \hat{\omega}^{-1/2} \rho_{x}^{\mu}
\right) + a\left(\overline{ \hat{\omega}^{-1/2}\rho_{x}^{\mu} } \right) \right).
\end{eqnarray}
By Nelson's analytic vector theorem, we can see that for each $x$, $A_{\mu}(x)$ is essentially self-adjoint on
$\mathcal{F}_{ \mathrm{fin} }.$
Here
$$\mathcal{F}_{ \mathrm{fin} }=\mathcal{L}(\,\{\Omega, a^{\dagger}(f_{1})\ldots a^{\dagger}(f_{n})\Omega |f_{j}\in L^{2}(\mathbb{R}^{3};
\mathbb{C}^{2}),j=1,\cdots,n,n\in\mathbb{N}\}\,).$$
The self-adjoint operator $A_{\mu}$ is defined by
\begin{eqnarray}
A_{\mu}=\int_{\mathbb{R}^{3}}^{\oplus}\overline{ A_{\mu}(x)|_{ \mathcal{F}_{ \mathrm{fin} } } }\,dx,\quad  A=(A_{1},A_{2},A_{3}).
\end{eqnarray}
\end{df}
\begin{df}[Free field Hamiltonian]
For $m\geq 0$, the free field Hamiltonian $\hfm $ is defined by the second quantization of  \[\hat{\omega}_{m}=\sqrt{-\Delta_{x} +v(x)+m^{2}}.\]
Namely
\begin{eqnarray}
\hfm =d\Gamma(\hat{\omega}_{m}).
\end{eqnarray}
\end{df}
Let $p=-i\nabla_{x}$.
Formally the Pauli-Fierz Hamiltonian with a variable mass is given by
\begin{eqnarray}
H_{m}^{V}\stackrel{\mathrm{formal}}{=}\frac{1}{2}\sum_{\mu,\nu}\left(p_{\mu}\otimes1 +\sqrt{\alpha}A_{\mu}\right)a_{\mu\nu}\left(p_{\nu}\otimes1 +\sqrt{\alpha}A_{\nu}\right)
+1\otimes \hfm +V\otimes 1.
\end{eqnarray}
Here $(a_{\mu\nu})_{\mu,\nu=1,2,3}=(a_{\mu\nu}(x))_{\mu,\nu=1,2,3}$ is positive definite for all $x\in\mathbb{R}^3$.
We consider only the case of $a_{\mu\nu}=\delta_{\mu,\nu}$ for simplicity in this paper.
We rigorously define the Pauli-Fierz Hamiltonian as a self-adjoint operator through a quadratic form.
\begin{df}
The quadratic form $q_{m}^{V}$ is defined by
\begin{eqnarray}
\lefteqn{q_{m}^{V}(\Psi,\Phi)=\frac{1}{2}\sum_{\mu=1}^{3}\left( (p_{\mu}\otimes 1 +\sqrt{\alpha} A_{\mu})\Psi,(p_{\mu}\otimes 1+\sqrt{\alpha}A_{\mu}) \Phi\right )}\nonumber \\
&&+\left( 1\otimes \hfm ^{1/2}\Psi,1\otimes \hfm ^{1/2}\Phi \right)
 +\left( V_{+}^{1/2}\otimes 1 \Psi,V_{+}^{1/2}\otimes 1\Phi \right) -\left( V_{-}^{1/2}\otimes 1\Psi,V_{-}^{1/2}\otimes 1\Phi \right) \nonumber\\
 \label{Hamiltonian}
\end{eqnarray}
with the form domain
\begin{eqnarray}
Q(q_{m}^{V})=D(|p|\otimes 1)\cap D(1\otimes \hfm ^{1/2})\cap D(|V|^{1/2}\otimes 1).
\end{eqnarray}
When $m=0$, we denote simply $q^{V}$ for $q_{0}^{V}$.
\end{df}
\subsection{Properties of the generalized eigenfunction $\Psi(k,x)$}
In this section we see regularity properties of $\Psi(k,x)$.
From \cite[Theorem 3]{Ik}, it follows that
\eq{50}
\sup_{k\in D,x\in\mathbb{R}^{3}}|\Psi(k,x)| <\infty
\en
 for any compact set $D$ but $0\notin D$.
Then by $(\ref{Lip-Sch})$, we see that
\begin{eqnarray}
\Psi(k,x)=e^{ikx}-\sum_{n=1}^{\infty}\left(\frac{1}{4\pi}\right)^{n}\int_{\mathbb{R}^{3n}} \frac{e^{i|k|\sum_{j=1}^{n}|y_{j}-y_{j-1}|}\,\Pi_{j=1}^{n}\,v(y_{j})}{ \Pi_{j=1}^{n} \, |y_{j}-y_{j-1}| } dy_{1}\cdots dy_{n} \label{PSI},
\end{eqnarray}
where $y_{0}=x$.
\begin{lem}\label{Psi}
Suppose Assumption \ref{v}. Then (a)--(c) hold.\\
(a)\cite[Lemma4.6 (3)]{GHPS} It holds that
\begin{eqnarray}
|\,\Psi(k,x)-e^{ikx}\,|\leq \mathrm{const}.\; \left<x\right>^{-1}. \label{Ps}
\end{eqnarray}
(b) $\Psi(k,x)$ is continuously differentiable in $x$ for each fixed $k$ but $k\neq 0,$ and
\begin{align}
\frac{\partial \Psi}{\partial x_{\mu}}(k,x)
=ik_{\mu}e^{ikx}-
\frac{1}{4\pi}\int_{\mathbb{R}^{3}} \left(\frac{1}{|x-y|^{3}}
-\frac{i|k|}{|x-y|^{2}}\right)
(x_{\mu}-y_{\mu})e^{i|k||x-y|}v(y)\Psi(k,y)dy. \label{Psi.1}
\end{align}
In particular, for any compact set $D$ but $0\notin D$,
\eq{51}
\sup_{k\in D ,x\in \mathbb{R}^{3}} \left| \frac{\partial \Psi}{\partial x_{\mu}}(k,x)\right|<\infty.
\en
(c) For $k\neq 0$, $h\neq 0$ and $k+h\neq 0$,
\begin{eqnarray}
\frac{1}{|h|}|\Psi(k+h,x)-\Psi(k,x)|&\leq& \mathrm{const.}(1+|x|),\label{PSI.1}\\
\frac{1}{|h|}\left| \frac{\partial\Psi}{\partial x_{\nu}}(k+h,x)-\frac{\partial\Psi}
{\partial  x_{\nu}}(k,x)\right|&\leq& \mathrm{const.}(1+|k|+|x|+|k||x|)
\label{PSI.2}
\end{eqnarray}
hold, and
$\Psi(k,x)$ and $\frac{\partial}{\partial x_{\nu}}\Psi(k,x)$ are differentiable in $k\in \mathbb{R}^{3}\setminus \{0\}$ for each fixed $x$.
\end{lem}
\begin{pf}
(a) Notice that for $0<a<3$ and $3<b$,
\begin{eqnarray}
\int_{\mathbb{R}^{3}}\frac{1}{|x-y|^{a}\left<y\right>^{b}}dy \leq \frac{4\pi}{3-a} \langle x\rangle^{-b}. 
\label{(23)}
\end{eqnarray}
Then (\ref{Ps}) follows from Assumption \ref{v} (1) and (\ref{PSI}).

(b) Let $k\neq 0$ be fixed.
Note that for $0<a<3$, $\frac{\chi_{\{|x|\leq 1\}}}{|x|^{a}}$ is integrable and $v(x)\Psi(k,x)$ is bounded. Since the convolution of an $L^{1}$-function and an $L^{\infty}$-function is continuous, we see that
$\int_{|x-y|\leq 1}\,\frac{e^{i|k||x-y|}v(y)}{|x-y|^{a}}\Psi(k,y) dy$
is continuous in $x$. Also by the dominated convergence theorem,
$\int_{|x-y|>1}\,\frac{e^{i|k||x-y|}v(y)}{|x-y|^{a}}\Psi(k,y) dy$
is continuous in $x$.
Thus by (\ref{Lip-Sch}), $\Psi(k,x)$ is continuous in $x$ for each fixed $k\neq 0$.
Let $u\in C_{c}^{\infty}(\mathbb{R}^{3})$.
We consider the integral:
\begin{eqnarray}
\int_{\mathbb{R}^{3}} \partial_{x,\mu} u(x)\left( \int_{\mathbb{R}^{3}} \frac{e^{i|k||x-y|}v(y)}{|x-y|}\Psi(k,y)dy\right)dx. \label{ap.1}
\end{eqnarray}
Since $|\partial_{x,\mu} u(x)|\frac{v(y)}{|x-y|}|\Psi(k,y)|$ is integrable in $x$ and $y$,
we use Fubini's theorem to see
\begin{eqnarray}
\text{(\ref{ap.1})}=\int_{\mathbb{R}^{3}} v(y)\Psi (k,y) \left(\int_{\mathbb{R}^{3}}(\partial_{x,\mu}u)(x)\frac{e^{i|k||x-y|}}{|x-y|}dx\right)dy.
\end{eqnarray}
Since for each $y$ and $k\neq 0$, the suface integral on the sphere with radius r centered at the origin
\begin{eqnarray}
\left|\int_{|x-y|=r} u(x)\frac{e^{i|k||x-y|}}{|x-y|}dS_{x}\right|
\end{eqnarray}
goes to $0$ as $r\to 0$, 
(\ref{ap.1}) can be written as
\begin{eqnarray}
-\int_{\mathbb{R}^{3}} v(y)\Psi (k,y) \left(\lim_{r\to 0}\int_{|x-y|>r} u(x) \partial_{x,\mu}\frac{e^{i|k||x-y|}}{|x-y|}dx\right)dy.
\label{27}
\end{eqnarray}
By using Fubini's theorem in (\ref{27}) again, for $k\neq 0$, we have
\begin{eqnarray}
\text{(\ref{ap.1})}=-\int_{\mathbb{R}^{3}}u(x)\left(\int_{\mathbb{R}^{3}}\partial_{x,\mu}\frac{e^{i|k||x-y|}v(y)}{|x-y|}\Psi(k,y)dy\right)dx.\label{pp}
\end{eqnarray}
Therefore by (\ref{Lip-Sch}) and (\ref{pp}), we can see that the distributional derivative of $\Psi(k,\cdot)$ is a function and the weak derivative of $\Psi(k,\cdot)$ is
\begin{eqnarray}
ik_{\mu}e^{ikx}-\frac{1}{4\pi}\int_{\mathbb{R}^{3}} \left(\frac{1}{|x-y|^{3}}
-\frac{i|k|}{|x-y|^{2}}\right)\,(x_{\mu}-y_{\mu})e^{i|k||x-y|}v(y)\Psi(k,y) dy,
\end{eqnarray}
which is continuous in $x$ for each $k\neq 0$ and bounded on $(k,x)\in D\times \mathbb{R}^{3}$.
Thus (b) follows.

(c)
Since
$
|e^{ikx}-1|\leq |k||x|
$,
we see that
\begin{eqnarray}
\lefteqn{\frac{1}{|h|} \, \left(\,\frac{1}{4\pi}\,\right)^{n}\,\left|\,
\frac{(e^{i|k+h|\sum_{j}|y_{j}-y_{j-1}|}-e^{i|k|\sum_{j}|y_{j}-y_{j-1}|})\Pi_{j}v(y_{j})}{\Pi_{j=1}^{n}\,|\,y_{j}-y_{j-1}\,|}
\right|}\nonumber \\
&&\leq \left(\,\frac{1}{4\pi}\,\right)^{n}\sum_{l=1}^{n}\,\frac{\Pi_{j=1}^{n}|\,v(y_{j})\,|}{\Pi_{j\neq l}\,|\,y_{j}-y_{j-1}\,|}.\hspace{3cm}
\end{eqnarray}
Moreover by (\ref{(23)}),
\begin{eqnarray}
\int \, \sum_{l=1}^{n}\,\frac{\Pi_{j=1}^{n}\,|\,v(y_{j})\,|}{\Pi_{j\neq l}\,|\,y_{j}-y_{j-1}\,|}dy_{1}\cdots dy_{n}<\mathrm{const.}\,n.
\end{eqnarray}
Thus by $(\ref{PSI})$ and by the Lebesgue dominated convergence theorem, (\ref{PSI.1}) follows and $\Psi(k,x)$ is differentiable in $k\in\mathbb{R}^{3}\setminus\{0\}$.
By means of (\ref{Psi.1}), we also see that (\ref{PSI.2}) holds and $\partial_{x,\nu}{\Psi}$ is differentiable in $k\in\mathbb{R}^{3}\setminus\{0\}$.\qed
\end{pf}
We here explain Lemma \ref{Psi}.
(a) is used in the localization estimate in Section 4.2,
(b) in Lemmas \ref{lem of p.t.} and \ref{number b'dd}, where
the pull through formula and a photon number bound are  derived.
(c) is used in Lemma \ref{derivative b'dd}, where a photon
derivative bound is shown.

Under Assumption \ref{v}, the generalized Fourier transformation on $L^{2}(\mathbb{R}^{3})$ is defined by
\begin{eqnarray}\label{4}
\mathscr{F} f(k)=(2\pi )^{-3/2}\mathrm{l.i.m.}\int f(x) \overline{\Psi (k,x)} dx, \label{Fourier}
\end{eqnarray}
which is unitary on $L^{2}(\mathbb{R}^{3})$ and its inverse transformation is given by
\begin{eqnarray}
\mathscr{F}^{-1} g(x)=(2\pi )^{-3/2}\mathrm{l.i.m.}\int g(k)\Psi (k,x) dk.
\end{eqnarray}
By the generalized Fourier transformation, $\hat{\omega}_{m}$ is transformed to the multiplication operator by
\begin{eqnarray}
\omega_{m}(k)=\sqrt{k^{2}+m^{2}}.
\end{eqnarray}
Namely
\begin{eqnarray}
\mathscr{F}\hat{\omega}_{m}\mathscr{F}^{-1}=\omega_{m}. \label{G.F.omega}
\end{eqnarray}
In what follows, we denote $\omega_{m=0}$ by $\omega$.
For each $x\in\mathbb{R}^{3}$, the quantized radiation field $A_{\mu}(x)$ is also transformed by the unitary operator $\Gamma\left(\mathscr{F}\right)$ to
\begin{eqnarray}
\hat{A}_{\mu}(x)=\left( a^{\dagger}\left( \overline{G_{\mu}(\cdot,x )}\right)
                                   +a\left( G_{\mu}(\cdot ,x) \right)\right),
\end{eqnarray}
where $G_{\mu}=(G_{\mu,1},G_{\mu,2})$ and
\begin{eqnarray}
G_{\mu,j}(k,x)=\frac{1}{\sqrt{2}}\frac{\hat{\varphi}_{j}^{\mu}(k)\Psi(k,x)}{\sqrt{\omega(k)}}. \label{G}
\end{eqnarray}
The free field Hamiltonian $\hfm $ is also transformed to
\begin{eqnarray}
\hhfm  =d\Gamma(\omega_{m}).
\end{eqnarray}
Then the quadratic form $q_{m}^{V}$ is transformed to
\begin{eqnarray}
\hat{q}_{m}^{V}(\Psi,\Phi)&=&\lefteqn{q_{m}^{V}((1\otimes \Gamma\left( \mathscr{F} \right) )\Psi,(1\otimes \Gamma\left( \mathscr{F} \right) )\Phi) }\nonumber \\
&=&\frac{1}{2}\sum_{\mu=1}^{3}\left((p_{\mu}\otimes 1+\sqrt{\alpha}\hat{A}_{\mu})\Psi,(p_{\mu}\otimes 1+\sqrt{\alpha}\hat{A}_{\mu})\Phi \right)\nonumber \\
&&+\left(1\otimes \hhfm^{1/2} \Psi,1\otimes\hhfm^{1/2} \Phi\right) +\left(V_{+}^{1/2}\Psi,V_{+}^{1/2}\Phi\right)-\left(V_{-}^{1/2}\Psi,
V_{-}^{1/2}\Phi\right)\nonumber\\
\end{eqnarray}
and the form domain to
\begin{eqnarray}
Q(\hat{q}_{m}^{V})=D(|p|\otimes 1)\cap D(1\otimes \hhfm  ^{1/2})\cap D(|V|^{1/2}\otimes 1).
\end{eqnarray}
Our first task is to show that quadratic form $\hat{q}_{m}^{V}$ is lower semibounded.
\begin{prop}\label{b'dd}
For all $\epsilon > 0$, there exist constants $0<C_{1}<1/2$ and $0<C_{2}$ so that for all $\Psi \in Q(\hat{q}_{m}^{V})$,
\begin{eqnarray}
\hspace{-0.5cm}
\frac{1}{2}\sum_{\mu=1}^{3}\left\Vert (p_{\mu}\otimes 1+\sqrt{\alpha} \hat{A}_{\mu})\Psi\right\Vert^{2}
+\epsilon \left\Vert (1\otimes \hhfm  ^{1/2})\Psi \right\Vert^{2}
\geq C_{1}\left\Vert (|p|\otimes 1)\Psi\right\Vert^{2}-C_{2}\left\Vert \Psi\right\Vert^{2}.
\end{eqnarray}
\end{prop}
\begin{pf}
For simplicity, we denote $p\otimes 1$ by $p$.
For each $0<\lambda<1$, we have
\begin{eqnarray*}
\lefteqn{\frac{1}{2}\sum_{\mu=1}^{3}\left\Vert\, (p_{\mu}+\sqrt{\alpha} \hat{A}_{\mu})\Psi\,\right\Vert^{2}}\\
&&=\left(\sqrt{\frac{\lambda}{2}}\Vert\, |p|\,\Psi\, \Vert- \sqrt{\frac{\alpha}{2\lambda}}\Vert\, |\hat{A}|\,\Psi \,\Vert\right)^{2} +\frac{1-\lambda}{2}\Vert \,|p|\,\Psi \,\Vert^{2}\\
&&\quad + \sqrt{\alpha}\left(\sum_{\mu=1}^{3}\Re (\,p_{\mu}\Psi,\hat{A}_{\mu}\Psi \,)+\Vert \,|p|\, \Psi \,\Vert \,\Vert\, |\hat{A}|\, \Psi\Vert
\right)+\frac{\alpha}{2} \left(1-\frac{1}{\lambda}\right) \Vert \,|\hat{A}|\,\Psi\, \Vert^{2}\\
&&\geq \frac{1-\lambda}{2}\Vert\, |p|\,\Psi \Vert^{2}+\frac{\alpha}{2} \left(1-\frac{1}{\lambda}\right) \Vert\, |\hat{A}|\,\Psi\, \Vert^{2}.
\end{eqnarray*}
Since by $(\ref{b'dd1})$ and $(\ref{b'dd2})$, $|\hat{A}|$ is relatively bounded with respect to $\hfm ^{1/2}$,
the proposition follows.\qed
\end{pf}
We introduce assumptions on external potential $V$.
\begin{assumption}\label{V}
\begin{enumerate}
\item  $-\infty<V(x)<\infty$ for almost every $x\in\mathbb{R}^{3}$;
\item $V_{-}$ is infinitesimally small with respect to $p^{2}$ in the sense of form:
for all $\epsilon > 0$, there exists a positive constant $C_{\epsilon}$ so that for $\Psi\in D(|p|)$,
\begin{eqnarray}
\Vert \,V_{-}^{1/2}\,\Psi\, \Vert^2 \leq \epsilon\,  \Vert \,|p| \Psi\, \Vert^{2}+C_{\epsilon} \Vert\, \Psi\, \Vert^{2}.
\end{eqnarray}
\end{enumerate}
\end{assumption}
\begin{prop}\label{p.a}
Suppose Assumptions \ref{v}, \ref{varphi} and \ref{V}.
Then for $m\geq 0$, $\hat{q}_{m}^{V}$ is symmetric, closed and bounded below.
\end{prop}
\begin{pf}
For simplicity, we denote $V\otimes 1$ by $V$ and $1\otimes \hhfm  $ by $\hhfm  $ and so on.
By Assumption \ref{V} (3), we see that $\hat{q}_{m}^{V}$ is symmetric.
By Proposition $\ref{b'dd}$, there exist constants $0<C_{1}<1$ and $0<C_{2}$ so that
\begin{eqnarray}
\hat{q}_{m}^{V}(\Psi,\,\Psi)\geq  \frac{C_{1}}{2}\sum_{\mu=1}^{3} \left\Vert\, p_{\mu}\,\Psi\,\right\Vert^{2}-C_{2}\Vert\,\Psi\,\Vert^{2}+\Vert\,V_{+}^{1/2}\,\Psi\,\Vert^{2}
-\Vert\,V_{-}^{1/2}\Psi\,\Vert^{2}
\end{eqnarray}
and
by Assumption \ref{V} (2),
\begin{eqnarray}
\hat{q}_{m}^{V}(\Psi,\,\Psi)\geq  \frac{C_{1}'}{2}\sum_{\mu=1}^{3} \left\Vert\, p_{\mu}\,\Psi\,\right\Vert^{2}-C_{2}'\Vert\,\Psi\,\Vert^{2}+\Vert\,V_{+}^{1/2}\Psi\,\Vert^{2}\label{abc}
\end{eqnarray}
for some $0<C_{1}',C_{2}'$ and for all $\Psi\in Q(\hat{q}_{m}^{V})$.
Thus the quadratic form $\hat{q}_{m}^{V}$ is bounded below.
Suppose that $\{\Psi_{i}\}_{i=1}^{\infty} \subset Q(\,\hat{q}_{m}^{V}\,)$ satisfies that
$\Psi_{i}\rightarrow \Psi$ and $\hat{q}_{m}^{V}(\Psi_{i}-\Psi_{j},\,\Psi_{i}-\Psi_{j})\rightarrow 0$ as $i$ and $j$ go to infinity.
Then by $(\ref{abc})$, $\{V_{+}^{1/2}\Psi_{i}\}_{i=1}^{\infty}$ is a Cauchy sequence.
Since $V_{+}$ is closed, $\Psi\in D(V_{+}^{1/2})$
and $V_{+}^{1/2}\Psi_{i}\rightarrow V_{+}^{1/2}\Psi$ as $i\to\infty$.
Similarly we can see that  $\Psi \in D(\hhfm  ^{1/2})\cap D(|p|)$
and
 $\hhfm  ^{1/2}\Psi_{i}
\rightarrow \hhfm  ^{1/2}\Psi$,
$|p|\Psi_{i}\rightarrow |p|\Psi$, $\hat{A}_{\mu}\,\Psi_{i}\rightarrow \hat{A}_{\mu}\Psi$ as $i\to\infty$.
Then $\Psi\in Q(\hat{q}_{m}^{V})$ and $\hat{q}_{m}^{V}$ is closed.\qed\\
\end{pf}
\begin{df}[The Pauli-Fierz Hamiltonian]
By Proposition $\ref{p.a}$, there exists the unique self-adjoint operator $\hat{H}_{m}^{V}$ such that $Q(\hat{q}_{m}^{V})=D(|\hat{H}_{m}^{V}|^{1/2})$ and for all $\Psi$, $\Phi\in Q(\hat{q}_{m}^{V})$,
$$\hat{q}_{m}^{V}(\,\Psi,\,\Phi\,)-E^{V}_{m}(\,\Psi\,,\Phi\,)=\left(\,(\hat{H}_{m}^{V}-E^{V}_{m})^{1/2}\,\Psi,\,(\hat{H}_{m}^{V}-E^{V}_{m})^{1/2}\,\Phi\,\right).$$
Here we denote the bottom of the spectrum of $\hat{H}_{m}^{V}$ by $E^{V}_{m}$.
i.e.,
\begin{eqnarray}
E^{V}_{m}=\inf_{\Psi\in Q(\hat{q}_{m}^{V}),\Vert \Psi \Vert = 1}\hat{q}_{m}^{V}(\Psi,\Psi)=\inf\sigma(\hat{H}_{m}^{V})
\end{eqnarray}
and $\sigma(\hat{H}_{m}^{V})$ is the spectrum of $\hat{H}_{m}^{V}$. $\hat{H}_{m}^{V}$ is called the Pauli-Fierz Hamiltonian.
\end{df}

We denote $\hat{H}_{m=0}^{V}$ by $\hat{H}^{V}$.
We define the inner product on $Q(\hat{q}_{m}^{V})$ by
\begin{eqnarray}
(\Psi,\Phi)_{+1}=\hat{q}_{m}^{V}(\Psi,\Phi)-E_{m}^{V}(\Psi,\Phi)+(\Psi,\Phi).
\end{eqnarray}
Then $(Q(\hat{q}_{m}^{V}),(\cdot,\cdot )_{+1})$ is a Hilbert space.
\section{Binding condition and main theorems}
\subsection{Binding condition}
We fix functions $\phi_{R}$ and $\tilde{\phi}_{R}$ on $\mathbb{R}^{3}$ with $R>0$ such that
\begin{eqnarray}
\phi,\;\tilde{\phi}\in C^{\infty}(\mathbb{R}^{3}),\quad
0\leq \phi \leq 1,\quad
\phi(x)=\left\{
\begin{array}{l}
1,\;\text{ if }\;|x|<1,\\
0,\;\text{ if }\;|x|>2,
\end{array}
\right.\quad
\phi(x)^{2}+\tilde{\phi}(x)^{2}=1.
\end{eqnarray}
We set
\begin{eqnarray}
\phi_{R}(x)=\phi(x/R),\quad\tilde{\phi}_{R}(x)=\tilde{\phi}(x/R). \label{phi}
\end{eqnarray}
We define $E^{V}_{R,m}$ by
\begin{eqnarray}
E^{V}_{R,m}=\inf_{\Psi\in Q(\hat{q}_{m}^{V}),(\tilde{\phi}_{R}\otimes 1)\Psi\neq 0}\,
\frac{\hat{q}_{m}^{V}(\,(\tilde{\phi}_{R}\otimes 1)\,\Psi,\,(\tilde{\phi}_{R}\otimes 1)\,\Psi)}{(\,(\tilde{\phi}_{R}\otimes 1)\,\Psi,\,(\tilde{\phi}_{R}\otimes 1)\,\Psi\,)}.
\end{eqnarray}
\begin{assumption}[Binding condition]\label{B.C.}
We suppose that for $m\geq 0$,
\begin{eqnarray}
E^{V}_{m}<\lim_{R\to\infty}\,E^{V}_{R,m}.
\end{eqnarray}
\end{assumption}
\begin{prop}
Suppose Assumptions \ref{v}, \ref{varphi} and \ref{V}. Also assume one of following conditions (a), (b) and (c) hold.
\begin{itemize}
\item (a) $\lim_{|x| \to \infty}V(x) =\infty $;
\item (b) $V(x)\leq 0$ and for sufficiently large $|x|$, $V(x)\leq -\frac{Z}{|x|^{\alpha}}$ with $0\leq \alpha<2$ and $\lim_{|x|\to \infty}\,V(x)=0$;
\item (c) $V(x)\leq 0$ and for sufficiently small $|x|$, $V(x)\leq -\frac{Z}{|x|^{\beta}}$ with $2<\beta$ and $\lim_{|x|\to \infty}\,V(x)=0$.
\end{itemize}
Then the binding condition is satisfied.
\end{prop}
\begin{pf}
It is shown similarly to \cite[Proposition 3.11.]{hi19} \qed
\end{pf}
\subsection{Main theorems}
In this section we state main theorems in this paper.
Let us define a useful dense subspace of $\mathcal H$,  $\mathcal S$,  by
$$\mathcal{S} =\mathcal{L}\left( \,\{f\otimes\Omega,\; f\otimes a^{\dagger}(f_{1}) \cdots a^{\dagger} (f_{n})\Omega \;|\; f,\;f_{j}\in C_{c}^{\infty}(\mathbb{R}^{3}),j=1,\cdots,n,\;n\in \mathbb{N} \}\,\right). $$
\begin{prop}\label{Core}
Let $V=0$. Suppose Assumptions
\ref{v}, \ref{varphi} and \ref{V}.
Then for $\Psi\in Q(\hat{q}_{m}^{0})$, there exists a sequence $\{\Psi^{j}\}_{j=1}^{\infty}\subset \mathcal{S}$ so that
$\Psi^{j}\rightarrow \Psi$ in $Q(\hat{q}_{m}^{0})$ as $j\to\infty$.
In particular, $\mathcal{S}$ is a form core for $\hat{q}_{m}^{0}.$
\end{prop}
\begin{pf}
Note that for $\Psi\in \mathcal{S}$,
\begin{eqnarray}
\hat{q}_{m}^{0}(\Psi,\,\Psi)\leq \sum_{\mu=1}^{3}\,(\,\Vert\, (p_{\mu}\otimes 1)\Psi \,\Vert^{2}+\alpha\Vert\, \hat{A}_{\mu}\Psi \,\Vert^{2}\,)+ \Vert\, (1\otimes \hhfm  ^{1/2})\,\Psi\, \Vert^{2} \nonumber \\
\leq \mathrm{const.}\,\left(\Psi,(|p|^{2}\otimes 1+1\otimes \hhfm  +1)\Psi\right).
\end{eqnarray}
Since $\mathcal{S}$ is an operator core for $p^{2}\otimes 1+1\otimes \hhfm  $
and $\hat{q}_{m}^{0}$ is closed, we can show the proposition.\qed
\end{pf}
Now we state main theorems.
\begin{thm}\label{main 1}
Suppose Assumption \ref{v}, \ref{V}, \ref{B.C.} and (1) in Assumption \ref{varphi} and $m>0$.
Then a ground state of $\hat{H}_m^{V}$ exists.
\end{thm}
For the case of $m=0$ to prove the existence of ground states, we introduce additional assumptions.
\begin{assumption}\label{as.exp}
There exist $Z$ and $W:\mathbb{R}^3\to\mathbb{R}$ such that (1)--(3) are satisfied.
\begin{enumerate}
\item $V=Z+W.$
\item $Z\in L^{1}_{\mathrm{loc}}(\mathbb{R}^{3})$ is bounded from below and (a) or (b) follows.
\begin{itemize}
  \item[(a)]  There exist   $\gamma>0$, $n\geq 1$ and a compact set $K$ such that
      $Z(x)\geq \gamma |x|^{2n}$ for $x\notin K$.
      \item[(b)] $\liminf_{|x|\to\infty}Z(x)>E^{V}_{0}$.
      \end{itemize}
\item \;$W\in L^{p}(\mathbb{R}^{3})$ with $p>\frac{3}{2}$ and $W<0$.
\end{enumerate}
\end{assumption}
Assumption \ref{as.exp} ensures a spatial exponential decay of a ground state of $\hat{H}_m^V$ uniformly in $m$.
\begin{thm}\label{main 2}
Let $m=0$ and suppose
Assumptions \ref{v}, \ref{varphi}, \ref{V}, \ref{B.C.} and \ref{as.exp}.
Then the ground state of $\hat{H}^{V}$ exists.
\end{thm}
\section{Massive case}
\subsection{Proof of Theorem \ref{main 1}}
\begin{lem}\label{m>0}
Let $\{ \Psi^{j}\}_{j=1}^{\infty}\subset Q(\hat{q}_{m}^{V})$ be any normalized sequence, converging weakly to zero in $\mathcal{H}$. Then
\begin{eqnarray}
\liminf_{j\to\infty}\,\hat{q}_{m}^{V}(\,\Psi^{j},\,\Psi^{j}\,)>E^{V}_{m}. \label{g}
\end{eqnarray}
\end{lem}
Before going to the proof of Lemma
\ref{m>0} we state a result immediately following from this lemma, and the proof will be given in Appendix A.
\begin{lem}\label{mgll}
Suppose that (\ref{g}) holds. Then Theorem \ref{main 1} is valid.
\end{lem}
{\it Proof of Lemma \ref{m>0} }
We can assume that $\{\Vert \Psi^{j} \Vert_{+1}\}_{j}$ is bounded.
Let $\phi_{R}$ and $\tilde{\phi}_{R}$ be functions in $(\ref{phi})$ and let $\Psi_{R}=(\phi_{R}\otimes 1)\Psi$ and $\tilde{\Psi}_{R}=(\tilde{\phi}_{R}\otimes 1)\Psi$.
Then the following identity follows:
\begin{eqnarray}
\hat{q}_{m}^{V}(\,\Psi^{j},\,\Psi^{j}\,)&=&\hat{q}_{m}^{V}(\,\Psi_{R}^{j},\,\Psi_{R}^{j}\,)+\hat{q}_{m}^{V}(\,\tilde{\Psi}_{R}^{j},\,\tilde{\Psi}_{R}^{j}\,) \nonumber \\
&&\qquad -\frac{1}{2}\Vert \,(|\nabla \phi_{R}|\otimes 1)\Psi^{j}\,\Vert^{2}-\frac{1}{2}\Vert \,(|\nabla \tilde{\phi}_{R}|\otimes 1)\Psi^{j}\,\Vert^{2}. \label{ac}
\end{eqnarray}
Take $j_{i}\in C^{\infty}(\mathbb{R}^{3})$, $i=1,2,$ such that
\begin{center}
$0\leq j_{i}(x) \leq 1,\quad \left\{
\begin{array}{cc}
j_{i}(x)=1, \text{ if } |x|<1 ,\\
j_{i}(x)=0,\text{ if }|x|>2
\end{array}
\right.$   and   $\;j_{1}^{2}+j_{2}^2=1.$
\end{center}
Put $j_{P,i}(k)=j_{i}(k/P)$ and $\hat{j}_{P,i}=j_{i}(-i \nabla_{k} /P)$.
Let $$U:\mathcal{F}(\,L^{2}(\mathbb{R}^{3};\mathbb{C}^{2})\oplus L^{2}(\mathbb{R}^{3};\mathbb{C}^{2})\,)\rightarrow \mathcal{F}(\, L^{2}(\mathbb{R}^{3};\mathbb{C}^{2})\,)\otimes \mathcal{F}(\,L^{2}(\mathbb{R}^{3};\mathbb{C}^{2})\,)$$ be the unitary operator
defined by
\begin{eqnarray*}
&&Ua^{\dagger}(f_{1}\oplus 0)\cdots a^{\dagger}(f_{k}\oplus 0)a^{\dagger}(0\oplus g_{1})\cdots a^{\dagger}(0\oplus g_{l})
\Omega\\
&&=a^{\dagger}(f_{1})\cdots a^{\dagger}(f_{k})\Omega\otimes a^{\dagger}(g_{1})\cdots a^{\dagger}(g_{l})\Omega.
\end{eqnarray*}
Let $\hat{j}_{P}:L^{2}(\mathbb{R}^{3};\mathbb{C}^{2})\rightarrow L^{2}(\mathbb{R}^{3}; \mathbb{C}^{2}) \oplus L^{2}(\mathbb{R}^{3};\mathbb{C}^{2})$ be the operator defined by
$$\hat{j}_{P}\Psi=\hat{j}_{P,1}\Psi\oplus \hat{j}_{P,2}\Psi.$$
Put $U_{P}=U\Gamma (\hat{j}_{P})$. Then $U_{P}$ is isometry from $\mathcal{F}$ to $\mathcal{F}\otimes\mathcal{F}$.
By (\ref{ac}), Lemmas \ref{locally.1}, \ref{locally.2}, \ref{locally.3} below and $\phi_{R}^{2}+\tilde{\phi}_{R}^{2}=1$, we have
\begin{align}
 &\liminf_{j\to\infty} \,\hat{q}_{m}^{V}(\Psi^{j},\Psi^{j}) \nonumber \\
&\geq\liminf_{j\to\infty}\,\left\{\,(\,E^{V}_{m}+m\,)\,\Vert \,\Psi_{R}^{j}\, \Vert^2
+E^{V}_{R,m}\,\Vert \,\tilde{\Psi}_{R}^{j}\, \Vert^{2}\right\}
+o(R^{0})+o_{R}(P^{0}) \nonumber \\
&=\liminf_{j\to\infty}\,\left\{ E^{V}_{m}
+m\,\Vert\,\Psi_{R}^{j}\,\Vert^{2}+(\,E^{V}_{R,m}-E^{V}_{m}\,)\,\Vert\,\tilde{\Psi}_{R}^{j}\,\Vert^{2} \right\}+o(R^{0})+o_{R}(P^{0})\nonumber\\
&\geq E^{V}_{m}+\min\{\,m,\,E^{V}_{R,m}-E^{V}_{m}\,\}+o(R^{0})+o_{R}(P^{0})\label{b'}
\end{align}
and then letting $P\to\infty$ in (\ref{b'}), we also have
\begin{eqnarray}
\liminf_{j\to\infty} \,\hat{q}_{m}^{V}(\Psi^{j},\Psi^{j})
\geq E^{V}_{m}+\min\{\,m,\,E^{V}_{R,m}-E^{V}_{m}\,\}+o(R^{0}).\label{b''}
\end{eqnarray}
Letting $R\to\infty$ again in (\ref{b''}), we have
\begin{eqnarray}
\liminf_{j\to\infty} \,\hat{q}_{m}^{V}(\Psi^{j},\Psi^{j})
\geq E^{V}_{m}+\min\{\,m,\,\lim_{R\to\infty}E^{V}_{R,m}-E^{V}_{m}\,\}.
\end{eqnarray}
Since $\min\{\,m,\,\lim_{R\to\infty}E^{V}_{R,m}-E^{V}_{m}\,\}>0$, Lemma \ref{m>0} is proven.\qed
\subsection{Localization estimates with variable mass}
The right hand side of (\ref{ac}) can be estimated in Lemmas \ref{locally.1} and \ref{locally.2} below separately.
\begin{lem}\label{locally.1}
It follows that
\begin{align}
\hat{q}_{m}^{V}(\,\Psi_{R}^{j},\,\Psi_{R}^{j}\,)
\geq
{E^{V}_{m}\Vert\,\Psi_{R}^{j}\,\Vert^{2}+m\Vert\,\Psi_{R}^{j}\,\Vert^{2}}
-\Vert\,(1\otimes (1\otimes P_{0}))\,(1\otimes U_{P})\,\Psi_{R}^{j}\,\Vert^{2}
+o_{R}(P^{0}). \label{u}
\end{align}
Here $P_{0}$ is the projection onto the subspace spanned by the Fock vacuum and $o_{R}(P^{0})$ is a function in $P$, which goes to zero as $P\to\infty$ for each fixed $R>0$.
\end{lem}
\begin{lem}\label{locally.2}
It follows that
\begin{eqnarray}
\hat{q}_{m}^{V}(\,\tilde{\Psi}_{R}^{j},\,\tilde{\Psi}_{R}^{j}\,)
\geq E^{V}_{R,m}\,\Vert \, \tilde{\Psi}_{R}^{j} \,\Vert^{2}+o(R^{0}). \label{b}
\end{eqnarray}
\end{lem}
Lemmas \ref{locally.1} and \ref{locally.2} will be proven in the next subsection.
We also use the lemma below, and the proof will be given in Appendix A.
\begin{lem}\label{locally.3}
As $j \to \infty$, it holds that
\begin{eqnarray}
\Vert \,(1_{\mathcal{H}}\otimes P_{0}))\,(1_{\mathcal{H}_{p}}\otimes U_{P})\,\Psi_{R}^{j} \,\Vert^2 \rightarrow 0. \label{e}
\end{eqnarray}
\end{lem}
It remains to prove Lemmas \ref{locally.1}, \ref{locally.2} and \ref{locally.3}.\\
{\it Proof of Lemma \ref{locally.1}.}
We will prove that the first term of the right hand side of (\ref{ac}) can be written as
\begin{eqnarray}
\hat{q}_{m}^{V}(\,\Psi_{R}^{j},\,\Psi_{R}^{j}\,)
=q^{V}_{m,R,P,1}(\,\Psi^{j},\,\Psi^{j}\,)+q^{0}_{m,R,P,2}(\,\Psi^{j},\,\Psi^{j}\,)+o_{R}(P^{0}),\label{a}
\end{eqnarray}
where
\begin{eqnarray}
\lefteqn{\!\!\!\!q_{m,R,P,1}^{V}(\Psi,\,\Phi)= }\nonumber\\
&&\!\!\!\!\!\!\!\!\!\!\!\!\!\!\!\frac{1}{2}\sum_{\mu=1}^{3}\left( (p_{\mu}\otimes 1_{\mathcal{F}\otimes \mathcal{F}} +\sqrt{\alpha}\hat{A}_{\mu}^{(1)})(1\otimes U_{P})\Psi_{R} ,(p_{\mu}\otimes 1_{\mathcal{F}\otimes \mathcal{F}}+\sqrt{\alpha}\hat{A}_{\mu}^{(1)})(1\otimes U_{P})\Phi_{R} \right) \nonumber\\
&&\quad+\left(\, (1\otimes \hhfm^{(1)1/2} \,)\,(1\otimes U_{P})\,\Psi_{R},\,(1\otimes \hhfm^{(1)1/2} \,)\,(1\otimes U_{P})\,\Phi_{R}\right) \nonumber\\
&&\quad+\left(\, (V_{+}^{1/2}\otimes 1_{\mathcal{F}\otimes \mathcal{F}})\,(1\otimes U_{P})\,\Psi_{R},\, (V_{+}^{1/2}\otimes 1_{\mathcal{F}\otimes \mathcal{F}})\, (1\otimes U_{P}) \,\Phi_{R} \, \right) \nonumber \\
&&\quad-\left(\, (V_{-}^{1/2}\otimes 1_{\mathcal{F}\otimes \mathcal{F}})\,(1\otimes U_{P})\,\Psi_{R},\, (V_{-}^{1/2}\otimes 1_{\mathcal{F}\otimes \mathcal{F}})\, (1\otimes U_{P}) \,\Phi_{R} \, \right)
\end{eqnarray}
and
\begin{eqnarray}
{ q_{m,R,P,2}^{0}\,(\,\Psi,\,\Phi\,)= }
 \left( \,(1\otimes \hhfm  ^{(2)1/2}\,)\,(1\otimes U_{P})\,\Psi_{R},\,(1\otimes \hhfm  ^{(2)1/2}\,)\,(1\otimes U_{P})\,\Phi_{R}\,\right).\qquad
\end{eqnarray}
Here
$
\hhfm  ^{(1)}=\hhfm  \otimes 1$,
$\hhfm  ^{(2)}=1\otimes \hhfm$,
\begin{eqnarray}
&&
\hat{A}^{(1)}=\int_{\mathbb{R}^{3}}^{\oplus} \, \overline{ \{\;(\;a^{\dagger}(\,\overline{G}(\cdot,x)\,)+a(\,G(\cdot,x)\,)\;)\otimes 1\;\} |_{ \mathcal{F}_{\mathrm{fin}} \hat{\otimes}\mathcal{F} }}\,dx,\\
&&\hat{A}^{(2)}=\int_{\mathbb{R}^{3}}^{\oplus} \, \overline{ \{\;1\otimes (\;a^{\dagger}(\,\overline{G}(\cdot,x)\,)+a(\,G(\cdot,x)\,)\;)\;\}|_{\mathcal{F}\hat{\otimes}\mathcal{F}_{\mathrm{fin}} }     }\,dx.
\end{eqnarray}
and
$\hat{\otimes}$ denotes the algebraic tensor product.
Noting $\hhfm  \geq m(1-P_{0})$, we can see that Lemma \ref{locally.1} follows from (\ref{a}) since
$$q_{m,R,P,1}^{V}(\Psi^{j},\,\Psi^{j})\geq E^{V}_{R,m}$$
and
$$q_{m,R,P,2}^{0}(\Psi_{j},\,\Psi_{j})\geq m\Vert\Psi^{j}_{R}\Vert^{2}-\Vert(1_{\mathcal{H}}\otimes P_{0})(1_{\mathcal{H}_{p}}\otimes U_{P})\Psi^{j}_{R}\Vert^{2}.$$
We divide the proof of (\ref{a}) into two steps.\\
{\bf Step 1}
\; We estimate the free field Hamiltonian part in (\ref{a}).
We denote $1\otimes T$ on $\mathcal{F}\otimes \mathcal{F}$ by $T$ for simplicity.
For $\Psi\in \mathcal{S}$,
\begin{eqnarray}
\lefteqn{(\,\Psi_{R}, \,\hhfm  \,\Psi_{R} \,)-
(\,\Psi_{R},\,U_{P}^{*}\;\left( \hhfm  ^{(1)} + \hhfm  ^{(2)} \right)\;
 U_{P}\,\Psi_{R}\,)}\nonumber \\
&=&(\,\Psi_{R},\, \hhfm  \,\Psi_{R}\, )
-(\,\Psi_{R},\,\Gamma(j_{P})^{*}
 d\Gamma(\omega_{m}\oplus\omega_{m})
\Gamma(j_{P})\;\Psi_{R})\nonumber\\
&=&(\,\Psi_{R},\,\hhfm  \,\Psi_{R}\,)-(\,\Psi_{R},\, (\,\hhfm  + d\Gamma(\;\hat{j}_{P,1}[\omega_{m},\hat{j}_{P,1}]\;)+ d\Gamma(\;\hat{j}_{P,2}[\omega_{m},\hat{j}_{P,2}]\;)\;)\,\Psi_{R})\nonumber \\
&=&-\left(\,\Psi_{R},\,\big(\;d\Gamma(\,\hat{j}_{P,1}[\omega_{m},\,\hat{j}_{P,1}]\,)+ d\Gamma(\,\hat{j}_{P,2} [\omega_{m},\,\hat{j}_{P,2}]\,)\;\big)\,\Psi_{R}\right).
\end{eqnarray}
By Proposition \ref{app.a} in Appendix B, we see that
\begin{eqnarray}
\lefteqn{\left|\, \Vert\, \hhfm  ^{1/2}\,\Psi_{R}\, \Vert^{2}-
\Vert \,(\,\hhfm  ^{(1)}+ \hhfm  ^{(2)}\,)^{1/2}\, U_{P}\,\Psi_{R}\,\Vert^{2}\right| }\nonumber\\
&&\leq (\,\Vert\, [\hat{j}_{P,1},\,\omega_{m}]\,\Vert+\Vert\, [\,\hat{j}_{P,2},\,\omega_{m}\,]\,\Vert )\;
(\,\Psi_{R},\,N\,\Psi_{R}\,)\nonumber \\
&&\leq \frac{C}{P}\Vert \,\hfm ^{1/2}\,\Psi_{R}\,\Vert^{2},
\label{aa}
\end{eqnarray}
where $C$ is a constant depending on $m$.
For $\Psi \in D(1\otimes\hhfm  ^{1/2})$, there exists a sequence $\{ \Psi^{j}\}_{j=1}^{\infty} \subset \mathcal{S} $ such that
$\Psi^{j}\rightarrow \Psi$ and $\hhfm  ^{1/2}\Psi^{j}\rightarrow \hhfm  ^{1/2}\Psi$ as $j\to\infty$
since $\mathcal{S}$ is a core of $\hhfm  ^{1/2}$.
Then $\Psi^{j}_{R}\rightarrow \Psi_{R}$ and $\hhfm  ^{1/2}\Psi^{j}_{R}\rightarrow \hhfm  ^{1/2}\Psi_{R}$ as $j\to\infty$.
Using (\ref{aa}) and the closedness of $\left(\, \hhfm  ^{(1)}+\hhfm  ^{(2)}\,\right)^{1/2}$,
we see that (\ref{aa}) holds for all $\Psi\in D(\hhfm  ^{1/2})$.
Therefore for all $\Psi\in Q(\hat{q}_{m}^{V})$, we have
\begin{eqnarray}
\left| \,\Vert \,\hhfm  ^{1/2}\,\Psi_{R} \,\Vert^{2}-
\Vert \,(\,\hhfm  ^{(1)}+\hhfm  ^{(2)}\,)^{1/2} \,U_{P}\,\Psi_{R}\,\Vert^{2}\, \right|
\leq \frac{C}{P}\, \Vert\, \Psi\, \Vert_{+1}^{2}.
\end{eqnarray}
{\bf Step 2}
 We shall estimate the kinetic energy part in (\ref{a}) and show that
\begin{align}
\Big| \left\Vert\, (\,p\otimes 1_{\mathcal{F}}\, +\sqrt{\alpha}\hat{A}\,)\,\Psi_{R}\,\right \Vert^{2}
-\left\Vert(p\otimes 1_{\mathcal{F} \otimes \mathcal{F}}+\sqrt{\alpha}\hat{A}^{(1)})(1\otimes U_{P})\Psi_{R}\right\Vert^{2}\Big|
=o_{R}(P^{0})\Vert\Psi\Vert_{+1}^{2}. \label{kin}
\end{align}
Let $Q$ be defined by
\begin{eqnarray}
Q=( 1\otimes U_{P} )\, (p\otimes 1_{\mathcal{F}}+\sqrt{\alpha}\hat{A})\,-\,(p\otimes 1_{\mathcal{F} \otimes \mathcal{F}}+\sqrt{\alpha}\hat{A}^{(1)})\, (1\otimes U_{P}).
\end{eqnarray}
Then
\begin{eqnarray}
\lefteqn{\left| \, \Vert\, (p\otimes 1_{\mathcal{F}} +\sqrt{\alpha}\hat{A})\,\Psi_{R}\, \Vert^{2}
-\Vert \,(\,p\otimes 1_{\mathcal{F} \otimes \mathcal{F}}+\sqrt{\alpha}\hat{A}^{(1)}\,)\,(1\otimes U_{P})\,\Psi_{R} \,\Vert^{2}\, \right| } \nonumber\\
&=&\Big|\, \left(\,(1\otimes U_{P})\,(p\otimes 1_{\mathcal{F}}+\sqrt{\alpha}\hat{A})\Psi_{R},Q\Psi_{R} \right)\nonumber \\
&&\quad+\left( Q\Psi_{R},(1\otimes U_{P})(p\otimes 1_{\mathcal{F}}+\sqrt{\alpha}\hat{A})\Psi_{R} \right)
-\left(Q\Psi_{R},Q\Psi_{R}\right)\, \Big|\nonumber \\
&\leq& 2\,\Vert\, (p\otimes 1_{\mathcal{F}}+\sqrt{\alpha}\hat{A})\,\Psi_{R}\,\Vert\, \Vert\, Q\,\Psi_{R}\, \Vert +\Vert\, Q\,\Psi_{R}\, \Vert^{2}\nonumber \\
&\leq& \mathrm{const.}\,(\, \Vert \,\Psi\, \Vert_{+1} +2\,\Vert\, (\nabla \phi_{R}\otimes 1_{\mathcal{F}})\,\Psi \Vert +\Vert \,Q\, \Psi_{R}\,\Vert\,)\,\Vert \,Q\, \Psi_{R}\,\Vert.
\end{eqnarray}
Thus it suffices to show that $\Vert\, Q\,\Psi_{R} \,\Vert = o_{R}(P^{0})\,\Vert \,\Psi\,\Vert_{+1}$.
Note that on $\mathcal{S}$,
\begin{eqnarray*}
\Gamma (\hat{j}_{P})\,a^{\dagger}(G)&=& a^{\dagger}(\hat{j}_{P}G)\Gamma(\hat{j}_{P}),\nonumber \\
\Gamma (\hat{j}_{P})\,a(\hat{j}_{P}^{*}G) &=& a(G)\,\Gamma (\hat{j}_{P}),\nonumber \\
U_{P}\,a^{\dagger}(G)&=&(\,a^{\dagger}(\hat{j}_{P,1} G)\otimes 1 + 1\otimes a^{\dagger}(\hat{j}_{P,2}G)\,)\,U_{P},\\
U_{P}\,a(\hat{j}_{P,1} G)&=&(a(G)\otimes 1)\,U_{P},\\
U_{P}\,a(\hat{j}_{P,2}G)&=&1\otimes a(G)\,U_{P}.
\end{eqnarray*}
Thus $Q$ can be written as
\begin{eqnarray}
Q=\sqrt{\alpha}\left\{-1\otimes \left(a^{\dagger}\left( ( 1-\hat{j}_{P,1} ) \overline{G} \right) \otimes 1\right) + 1\otimes \left(1\otimes a^{\dagger}\left( \hat{j}_{P,2} \overline{G} \right)\,\right)\,\right\}(1\otimes U_{P})\nonumber\\
+\sqrt{\alpha}\,(1\otimes U_{P})\,\left(1\otimes a\left((1-\hat{j}_{P,1})G\,\right)\,\right)
\end{eqnarray}
on $\mathcal{S}$.
Therefore
\begin{eqnarray}
\Vert \, Q\,\Psi_{R}\, \Vert
&\leq& \sqrt{\alpha}\sup_{x\in \mathbb{R}^{3}}\left(\Vert \,(\,1-\hat{j}_{P,1}\,)G(\cdot ,x) \,\Vert\; |\,\phi_{R}(x)\,|\right)\nonumber \\
&&\qquad\qquad\qquad (\,(1\otimes U_{P})\,\Psi,\,(1\otimes (N+1)\otimes 1)\,(1\otimes U_{P})\,\Psi\,)^{1/2}\nonumber \\
&&+\sqrt{\alpha}\sup_{x\in \mathbb{R}^{3}} \left(\Vert \, \hat{j}_{P,2}G(\cdot ,x) \, \Vert \;|\,\phi_{R}(x)\,|\right)\nonumber \\
&&\qquad\qquad\qquad (\,(1\otimes U_{P})\,\Psi,\,(1\otimes 1\otimes (N+1))\,(1\otimes U_{P})\,\Psi\,)^{1/2}\nonumber \\
&&+\sqrt{\alpha}\,\sup_{x\in \mathbb{R}^{3}} \left(\,\Vert (1-\hat{j}_{P,1})G(\cdot ,x) \Vert \;|\phi_{R}(x)|\,\right)\; (\,\Psi,\,(1\otimes N)\, \Psi\,)^{1/2}.
\end{eqnarray}
Note that
\[ U^{*}\, ((N+1) \otimes 1)\,)\, U \leq  U^{*}\Big((N+1) \otimes 1 + 1\otimes (N+1) \Big)\, U= N+1.\]
Then
\begin{eqnarray}
\Vert \,Q\, \Psi_{R}\, \Vert\leq 2\sqrt{\alpha}\,\sup_{x\in \mathbb{R}^{3}}\,\left(\Vert\, (\,1-\hat{j}_{P,1}\,)G(\cdot ,x) \,\Vert\, |\phi_{R}(x)|\right)\, \Vert\,(1\otimes (N+1))^{1/2}\, \Psi\,\Vert\nonumber \\
+\sqrt{\alpha}\sup_{x\in \mathbb{R}^{3}}\left(\Vert\, \hat{j}_{P,2}\,G(\cdot,x) \Vert |\,\phi_{R}(x)\,|\right) \Vert \,(1\otimes (N+1)\,)^{1/2}\, \Psi\, \Vert.\label{(94)}
\end{eqnarray}
We denote the Fourier transformation of $f\in L^{2}$ by $F[f]$. Then
\begin{eqnarray}
\sup_{x\in \mathbb{R}^{3}}\,\left(\Vert \,(1-\hat{j}_{P,1})\,G(\cdot ,x) \,\Vert \; |\,\phi_{R}(x)\,|\right)
&&=\sup_{|x|\leq R} \,\Vert \,(\,1-j_{P,1}\,)\,F[\,G(\cdot ,x)\,] \,\Vert \nonumber \\
&&\leq \sup_{|x|\leq R} g_{P}(x),\label{dini.1}
\end{eqnarray}
where
$$g_{P}(x)=\left(\sum_{j,\mu}\,\int_{|\xi|>P}\,\left|\,F\left[\,\frac{\hat{\varphi}_{j}^{\mu}\Psi(\cdot,x)}{\sqrt{2\omega}}\,\right] (\xi)\,\right|^{2}\,d\xi \right)^{1/2}.$$
For each $x$, $g_{P}(x)$ is monotonically converges to $0$ as $P\uparrow \infty$.
In addition,
\begin{eqnarray}
|g_{P}(x+h)-g_{P}(x)|\!&\leq&\! \Vert \,F[\,G(\cdot,x+h)-G(\cdot,x)\,] \,\Vert_{L^{2}} \nonumber \\
\!&=&\!\!\left(\,\int_{\mathbb{R}^{3}}\sum_{j,\mu} \left| \frac{\hat{\varphi}_{j}^{\mu} (k)}{\sqrt{2\omega(k)}}\right|^{2}|\Psi(k,x+h)-\Psi(k,x)|^{2}dk \right)^{1/2}\hspace{-0.5cm}.
\end{eqnarray}
Thus since $\sup_{k,x}|\Psi(k,x)|<\infty$ and $\Psi(k,x)$ is continuous in $x$ for each $k$ but $k\neq 0$, we see that
$g_{P}(x)$ is continuous in $x$ for each $P$.
By Dini's theorem, we have
\begin{eqnarray}
\sup_{|x|\leq R}\, |\phi_{R}(x)|g_{P}(x)
\rightarrow 0 \text{ as } P\to \infty.\label{dini.2}
\end{eqnarray}
Therefore by (\ref{dini.1}) and (\ref{dini.2}),
\begin{eqnarray}
\sup_{x\in \mathbb{R}^{3}}\,\left(\Vert \,(1-\hat{j}_{P,1})\,G(\cdot ,x) \,\Vert \; |\,\phi_{R}(x)\,|\right)\rightarrow 0 \text{ as } P\to \infty \label{(98)}
\end{eqnarray}
follows.
Similarly, we can see that
\begin{eqnarray}
\sup_{x\in \mathbb{R}^{3}}\,\left(\Vert \, \hat{j}_{P,2}\,G(\cdot ,x) \, \Vert\,|\phi_{R}(x)|\right) \rightarrow 0 \text{  as } P\to \infty. \label{(99)}
\end{eqnarray}
Thus by (\ref{(94)}), (\ref{(98)}) and (\ref{(99)}), we have
\begin{eqnarray}
\Vert \,Q\,(\phi_{R}\otimes 1)\, \Psi\, \Vert\ = o_{R}(P^{0})\,\Vert\, \Psi \,\Vert_{+1}.
\end{eqnarray}
Therefore (\ref{kin}) follows for $\Psi\in\mathcal{S}$.
In a similar argument to the last part of Step 1, we find that $(\ref{kin})$ holds for all $\Psi\in Q(\hat{q}_{m}^{V})$.
\qed\\
{\it Proof of Lemma \ref{locally.2}.}
 Let
\begin{eqnarray}
g_{P',1} (x,y) = \chi \left( \frac{x-y}{P'} \right)\;\text{  and  }\; g_{P',2}(x,y)=1-g_{P',1}(x,y),
\end{eqnarray}
where $\chi\in C_{c}^{\infty}(\mathbb{R}^{3}),\;0\leq \chi(x) \leq 1,\;
\left\{
\begin{array}{c}
\chi (x)=1\text{ if }|x|\leq 1,\\
\chi (x)=0\text{ if }|x|\geq 2.
\end{array}\right.$
Set
\begin{eqnarray*}
j_{P',i,x}'(y)&=&\frac{g_{P',i}(x,y)}{\sqrt{g_{P',1}(x,y)^{2}+g_{P',2}(x,y)^{2}}},\\
\hat{j}_{P',i,x}'&=&j_{P',i,x}'(-i\nabla),\;\, i=1,2,\\
\hat{j}_{P',x}'&=& \hat{j}_{P,1,x}'\,\oplus \,\hat{j}_{P,2,x}',\\
U_{P'}'(x)&=&U\Gamma(\hat{j}_{P,x}')\\
(U_{P'}'\Psi)(x)&=&U_{P'}'(x)\Psi(x).
\end{eqnarray*}
Similarly to the proof of Lemma \ref{locally.1}, we have
\begin{eqnarray}
&&{\Big|\, \left\Vert\, (\,1\otimes\hhfm  ^{1/2}\,)\,\tilde{\Psi}_{R} \,\right\Vert^{2} }
-\left \Vert\, \left(\,1\otimes \hhfm  ^{(1)}\,+1\otimes \hhfm  ^{(2)}\,\right)^{1/2} U_{P'}' \tilde{\Psi}_{R}\,\right \Vert^{2}\,\Big| \nonumber\\
&&\qquad\qquad\qquad\leq \frac{\mathrm{const.}}{P'}\Vert\, (\,1\otimes \hfm ^{1/2}\,)\tilde{\Psi}_{R}\,\Vert^{2}.
\end{eqnarray}
Let
\begin{eqnarray}
Q'=U_{P'}'\,(\,p\otimes 1_{\mathcal{F}}+\sqrt{\alpha}\hat{A}\,)\,-\,(\,p\otimes 1_{\mathcal{F}\otimes\mathcal{F}}+\sqrt{\alpha}\hat{A}^{(2)}\,)\,U_{P'}'.
\end{eqnarray}
Then
\begin{eqnarray}
\lefteqn{\Big| \,\left\Vert\, (\,p\otimes 1_{\mathcal{F}} +\sqrt{\alpha}\hat{A}\,)\tilde{\Psi}_{R}\,\right\Vert^{2}-\,\left\Vert\, p\otimes 1_{\mathcal{F}\otimes\mathcal{F}}+\sqrt{\alpha}\hat{A}^{(2)}\,)U_{P'}'\tilde{\Psi}_{R}\,\right\Vert^{2} \,\Big| } \nonumber \\
&&\leq
\mathrm{const.}\,\left( \Vert\, \Psi \,\Vert_{+1} +2\,\Vert\, (\nabla \tilde{\phi}_{R}\otimes 1_{\mathcal{F}})\,\Psi \,\Vert \,+\,\Vert\, Q' \tilde{\Psi}_{R}\,\Vert\,\right)\,\Vert \,Q'\tilde{\Psi}_{R}\, \Vert
\end{eqnarray}
and
\begin{eqnarray}
\lefteqn{Q'=U_{P'}'\,(\,p\otimes 1_{\mathcal{F}}\,)- (\,p\otimes 1_{\mathcal{F}\otimes\mathcal{F}} \,)\, U_{P'}'} \nonumber \\
&&\quad+\sqrt{\alpha}\left\{ -1\otimes a^{\dagger}\left(\; (\, 1-\hat{j}_{P',1,x}' \,)\,\overline{G}\;\right) \otimes 1 +1\otimes \left(1\otimes a^{\dagger}\left(\; \hat{j}_{P',2,x}'\,\overline{G}\;\right)\,\right)\right\}\,U_{P'}'\nonumber\\
&&\quad+\sqrt{\alpha}\,U_{P'}'\;\left(1\otimes a\left(\;(\,1-\hat{j}_{P,1}\,)\,G\;\right)\,\right)
\end{eqnarray}
on $\mathcal{S}$.
Thus for $\Psi\in \mathcal{S}$,
\begin{eqnarray}
{\Vert \, Q'\tilde{\Psi}_{R}\,\Vert} &\leq&\,\Vert \, U_{P'}'\,(\,p\otimes 1_{\mathcal{F}}\,)\tilde{\Psi}_{R}- (\,p\otimes 1_{\mathcal{F}\otimes\mathcal{F}} \,)\, U_{P'}'\, \tilde{\Psi}_{R}\,\Vert \nonumber \\
&&\quad+\sqrt{\alpha}\,\sup_{x}(\Vert \,(\,\hat{j}_{1,P',x}'-1\,)\,G\,\Vert\,|\tilde{\phi}_{R}(x)|\,)\,\left(\,\Psi,\, (1\otimes (N+1)\,)\,\Psi\,\right)^{1/2}\nonumber \\
&&\quad+2\sqrt{\alpha}\,\sup_{x}(\Vert \,   \hat{j}_{2,P',x}'\,G\,\Vert\,|\tilde{\phi}_{R}(x)|\,)\,\left(\,\Psi,\, (1\otimes (N+1)\,)\,\Psi\,\right)^{1/2}. \label{c}
\end{eqnarray}
We estimate the right hand side of (\ref{c}).
Let $$d\Gamma(A,B)=0\oplus \left[\,\oplus_{n=1}^{\infty}\overline{\sum_{k=1}^{n}A\otimes \cdots \otimes A \otimes \stackrel{kth}{B}\otimes A\otimes\cdots \otimes A}\,\right].$$
The first term of the right hand side of (\ref{c}) is
\begin{eqnarray}
{\Vert \, U_{P'}'\,(\,p\otimes 1_{\mathcal{F}}\,)\tilde{\Psi}_{R}- (\,p\otimes 1_{\mathcal{F}\otimes\mathcal{F}} \,)\, U_{P'}'\, \tilde{\Psi}_{R}\,\Vert}\;
&=&\Vert \, d\Gamma(\,\hat{j}_{P',x}',\;(-i\nabla_{x} \hat{j}_{P',x}')\,)\tilde{\Psi}_{R}\,\Vert\nonumber \\
&\leq& \frac{\mathrm{const.}}{P'}\Vert\, (1\otimes N^{1/2}) \,\tilde{\Psi}_{R}\, \Vert\nonumber \\
&\leq&\,\frac{\mathrm{const.}}{P'm^{1/2}}\Vert \,\Psi \,\Vert_{+1}.\label{loc.3}
\end{eqnarray}
Next we estimate the second term of the right hand side of (\ref{c}).
Note that $|y-x|\geq P'$ if $j_{1,P',x}'(y)\neq 1$. 
Thus by $(\ref{Ps})$,
\begin{align}
&\Vert \,(\,\hat{j}_{1,P',x}'-1)\,G\,\Vert \leq
\left(\sum_{\mu,j} \,\int_{|\xi-x|\geq P'} \,\left|\; F\left[\,\frac{\hat{\varphi}_{j}^{\mu}\,}{\sqrt{2\omega}}\,\Psi(\cdot,x)\,\right] (\xi)\;\right|^{2}\,d\xi \,\right)^{1/2} \nonumber \\
&\leq
\, 2\sum_{\mu,j} \,\left\Vert\, \frac{\hat{\varphi}_{j}^{\mu}}{\sqrt{2\omega}}\,\right\Vert \, \left\Vert \,\Psi(\cdot , x)- e^{ikx} \,\right\Vert_{L^{\infty}}+\left(\sum_{\mu,j}\,\int_{|\xi - x|\geq P'} \left| F\left[\frac{\hat{\varphi}_{j}^{\mu}}{\sqrt{2\omega}}\right](\xi -x)\right|^{2} d \xi \right)^{1/2} \nonumber \\
&\leq \,\mathrm{const.}\, \left<\,x\,\right>^{-1}\,+\,o(P'^{0}),
\end{align}
where we used $\hat{\varphi}_{j}^{\mu}/\sqrt{2\omega} \in L^{2}(\mathbb{R}^{3})$.
Therefore
\begin{eqnarray}
\sup_{x}\,\Vert \,(\,\hat{j}_{1,P',x}'-1)\,G\,\Vert\,|\tilde{\phi}_{R}(x)|\leq \mathrm{const.}\,\left<\,R\,\right>^{-1}+\,o(P'^{0}).\label{loc.1}
\end{eqnarray}
Similarly
\begin{eqnarray}
\sup_{x}\,\Vert \, \hat{j}_{P,2,x} \,G\,\Vert\,\leq\,\mathrm{const.}  \, \left<\,R\,\right>^{-1}+o(P'^{0}) \label{loc.2}
\end{eqnarray}
follows.
By (\ref{c}), (\ref{loc.3}), (\ref{loc.1}) and (\ref{loc.2}), we obtain that
\begin{eqnarray}
\Vert\, Q'\tilde{\Psi}_{R}\,\Vert =( o(R^{0})+o(P'^{0})\;)\Vert\Psi\Vert_{+1}.
\end{eqnarray}
Consequently, we have
\begin{eqnarray}
{\hat{q}_{m}^{V}(\tilde{\Psi}_{R},\tilde{\Psi}_{R})}
=\!\left\Vert(1\otimes \hhfm  ^{(1)1/2} )\!(1\otimes U_{P'}')\tilde{\Psi}_{R}\right\Vert^{2}\!+\!\tilde{q}_{m,P',R}^{V}(\Psi,\Psi)\!+\!o(R^{0})\!+\!o(P'^{0}),\label{d}
\end{eqnarray}
where
\begin{eqnarray}
\lefteqn{ \tilde{q}_{m,P',R}^{V}(\Psi,\Phi)=}\nonumber \\
&&\frac{1}{2}\!\!\left((p\otimes 1_{\mathcal{F}\otimes\mathcal{F}}+\sqrt{\alpha}\hat{A}^{(2)}) (1\otimes U_{P'}')\tilde{\Psi}_{R},(p\otimes 1_{\mathcal{F}\otimes\mathcal{F}}+\sqrt{\alpha}\hat{A}^{(2)}) (1\otimes U_{P'}')\tilde{\Phi}_{R} \!\right)\nonumber \\
&&+\left( (1\otimes \hhfm  ^{(2)1/2}) (1\otimes U_{P'}')\tilde{\Psi}_{R},(1\otimes \hhfm  ^{(2)1/2}) (1\otimes U_{P'}')\tilde{\Phi}_{R} \right)\nonumber \\
&&+\left( (V_{+}^{1/2}\otimes 1_{\mathcal{F}\otimes\mathcal{F}})(1\otimes U_{P'}')\tilde{\Psi}_{R}, (V_{+}^{1/2}\otimes 1_{\mathcal{F}\otimes\mathcal{F}})(1\otimes U_{P'}')\tilde{\Phi}_{R} \right)\nonumber \\
&&-\left( (V_{-}^{1/2}\otimes 1_{\mathcal{F}\otimes\mathcal{F}})(1\otimes U_{P'}')\tilde{\Psi}_{R}, (V_{-}^{1/2}\otimes 1_{\mathcal{F}\otimes\mathcal{F}})(1\otimes U_{P'}')\tilde{\Phi}_{R} \right).
\end{eqnarray}
Therefore (\ref{b}) holds since $\tilde{q}_{m,P',R}^{V}(\Psi,\Psi)\geq E^{V}_{R,m}$ and the left hand side is independent
of $P'$.
\qed\\
\section{Massless case}
In this section we assume Assumptions \ref{v},\ref{varphi},\ref{V} and
\ref{B.C.} with $m=0$ throughout. Let $\Phi_{m}$ be a normalized ground state of $\hat{H}_{m}^{V}$.
\subsection{Spatial exponential decay}
Lemma \ref{bbb} 
 will be proven similarly to \cite{GLL}. See also Appendix A.
\begin{lem}\label{bbb}
\begin{enumerate}
\item
$
\lim_{m\to 0} E^{V}_{m}=E^{V}_{0}.
$
\item 
For sufficiently small $m$, the binding condition holds.
\end{enumerate}
\end{lem}
A spatial exponential decay of a ground state of $\hat H_m^V$ is proven in \cite{FH}.
\begin{prop} \label{exp}\cite{FH}
Suppose Assumption \ref{as.exp}.
Then there exists $m_0>0$ such that
\begin{eqnarray}
\sup_{0<m<m_0} \Vert \exp (c|x|)\otimes 1 \,\Phi_{m} \Vert^{2}\leq C,
\end{eqnarray}
where $c$ and $C$ are constants independent of $m\in (0,m_0)$.
\end{prop}
\subsection{Boson number bound}
\begin{lem}
[Pull through formula]
\label{lem of p.t.}
Let $f \in D(\omega_{m})$. Then  $a(f)\Phi_{m}\in Q(\hat{q}_{m}^{V})$ and for all $\eta\in Q(\hat{q}_{m}^{V})$, we have
\begin{eqnarray}
&&
{\hat{q}_{m}^{V}(\,\eta,\,a(f)\,\Phi_{m}\,)-E^{V}_{m}(\,\eta,\,a(f)\,\Phi_{m}\,)}\nonumber \\
&&=-\sqrt{\alpha}\,(\,\eta,\,(\overline{f},\,\overline{G})\cdot (\,p+\sqrt{\alpha}\hat{A}\,)\,\Phi_{m}\,)
+\frac{i\sqrt{\alpha}}{2}(\eta,(\overline{f},\nabla_{x}\cdot \overline{G})\Phi_{m})
-(\,\eta,a(\omega_{m}f)\,\Phi_{m}\,).\nonumber\\
&&\label{s.8}
\end{eqnarray}
Here $G_{\mu,j}(k,x)=\frac{1}{\sqrt{2}}\frac{\hat{\varphi}_{j}^{\mu}(k)\Psi(k,x)}{\sqrt{\omega(k)}}.$
\end{lem}
\begin{pf}
For simplicity, we denote $A\otimes 1$ (resp. $1\otimes B$) by $A$ (resp. $B$).
Let
\begin{eqnarray}
a_{R}(f)&=&R(\,\hhfm  +R\,)^{-1}a(f), \nonumber \\
a_{R}^{\dagger}(f)&=&Ra^{\dagger}(f) ( \,\hhfm  +R\,)^{-1}
\end{eqnarray}
for all $R>0$.
We first show  that $a_{R}(f)$  and  $a_{R}^{\dagger}(f)$ are bounded operators on $Q(\hat{q}_{m}^{V})$; that is, there exists some constant $C(R)$ so that
\begin{eqnarray}
\Vert \,a_{R}(f)\,\Psi \Vert_{+1}&\leq& C(R)\,\Vert\, \Psi\, \Vert_{+1},\label{t.1} \\
\Vert\, a_{R}^{\dagger}(f)\,\Psi\, \Vert_{+1}&\leq& C(R)\,\Vert \,\Psi \,\Vert_{+1} \label{t.2}
\end{eqnarray}
hold for all $\Psi\in Q(\hat{q}_{m}^{V})$.
Let us first show (\ref{t.1}). It suffices to prove that
\begin{eqnarray}
R (\,\hat{H}_{m}^{V}-E^{V}_{m}+1\,)^{1/2}\,(\, \hhfm  +R\,)^{-1}\,a(f)\,(\,\hat{H}_{m}^{V}-E^{V}_{m}+1\,)^{-1/2}
\end{eqnarray}
is bounded on $\mathcal{H}$.
Let $\Psi\in D(\,|\hat{H}_{m}^{V}|^{1/2}\,)$ and put $$\Phi=(\,\hhfm  +R\,)^{-1}\,a(f)\,(\,\hat{H}_{m}^{V}-E^{V}_{m}+1\,)^{-1/2}\, \Psi.$$
Note that
\begin{eqnarray*}
{\Vert\,(\,\hat{H}_{m}^{V}-E^{V}_{m}+1\,)^{1/2}\Phi\,\Vert}
&\leq& \textrm{const.}\Vert \,(p^{2}+\hhfm  +V_{+}+1)^{1/2}\Phi\,\Vert\\
&\leq& \textrm{const.}(\,\Vert \,(p^{2}+V_{+}+1)^{1/2}\Phi\Vert+\Vert\,\hhfm  ^{1/2}\Phi\,\Vert\,).
\end{eqnarray*}
Also note that by (\ref{b'dd2}),
$$ \Vert\,(\hhfm  +R)^{-1/2}a(f)\,\Vert=\Vert\,a^{\dagger}(f)(\hhfm  +R)^{-1/2}\,\Vert<\infty $$
and since $p^{2}+V_{+}+1\leq \textrm{const.}\, (\,\hat{H}_{m}^{V}-E^{V}_{m}+1\,)$,
$$\Vert \,(\,p^{2}+V_{+}+1\,)^{1/2}\, (\,\hat{H}_{m}^{V}-E^{V}_{m}+1\,)^{-1/2} \Vert < \infty. $$
Then we see that
\begin{eqnarray}
\lefteqn{R\,\Vert\, (\,\hat{H}_{m}^{V}-E^{V}_{m}+1\,)^{1/2}\,(\,\hhfm  +R\,)^{-1}\,a(f)\,(\,\hat{H}_{m}^{V}-E^{V}_{m}+1\,)^{-1/2}\, \Psi\, \Vert }\nonumber \\
&&\leq \mathrm{const.}R\Big(\Vert(\hfm +R)^{-1}a(f)\Vert\,\Vert(p^{2}+V_{+}+1)^{1/2} (\hat{H}_{m}^{V}-E^{V}_{m}+1\,)^{-1/2}\Vert\nonumber \\
&&\quad+\,\Vert\, ( \,\hhfm  +R\,)^{-1/2} \, a(f) \,\Vert \;\Vert\,(\,\hat{H}_{m}^{V}-E^{V}_{m}+1)^{-1/2}\,\Vert\, \Big) \Vert\Psi\Vert.
\end{eqnarray}
Thus (\ref{t.1}) holds. Next we shall show (\ref{t.2}).
Since
\[ [a^{\dagger}(f), \,(\,\hhfm  +R\,)^{-1}]=(\,\hhfm  +R\,)^{-1}a^{\dagger}(\omega_{m}f)\,(\,\hhfm  +R\,)^{-1}\]
 on $\mathcal{S}$, we see that $\hhfm  ^{1/2} \,a^{\dagger}(f) \,(\,\hhfm  +R\,)^{-1}$ is bounded and
\begin{eqnarray}
{\Vert \,\hhfm  ^{1/2} \,a^{\dagger}(f) \,(\,\hhfm  +R\,)^{-1}\,\Vert} 
&\leq&  \Vert \,\hhfm  ^{1/2}(\hhfm  +R)^{-1} \,a^{\dagger}(f) \,\Vert \nonumber \\
&&+ \Vert \,\hhfm  ^{1/2}\,(\, \hhfm  +R\,)^{-1} \,a^{\dagger}(\omega_{m} f) \,(\, \hhfm  +1\,)^{-1} \,\Vert \nonumber \\
&\leq&  \Vert \,a(f)\,(\, \hhfm  +R\,)^{-1/2} \, \Vert \nonumber \\
&& + \Vert \,(\, \hhfm  +R\,)^{-1/2}\,\Vert \,\Vert\,a^{\dagger}(\omega_{m} f) \,(\, \hhfm  +R\,)^{-1} \,\Vert<\infty. \nonumber
\end{eqnarray}
Similarly to the proof of (\ref{t.1}), we have
\begin{eqnarray}
\lefteqn{R\,\Vert\,(\,\hat{H}_{m}^{V}-E^{V}_{m}+1)^{1/2}a^{\dagger}(f)(\,\hhfm  +R)^{-1}\,(\,\hat{H}_{m}^{V}-E^{V}_{m}+1\,)^{-1/2}\Psi \Vert} \nonumber \\
&&\leq \mathrm{const.}\,R\Big(\,
\Vert \,\hhfm  ^{1/2} \,a^{\dagger}(f)\, (\,\hhfm  +R\,)^{-1}\Vert\,\Vert\,(\,\hat{H}_{m}^{V}-E^{V}_{m}+1\,)^{-1/2}\,\Vert\nonumber \\
&&\quad+ \Vert \,(\,p^{2}+V_{+}+1\,)^{1/2}(\,\hat{H}_{m}^{V}-E^{V}_{m}+1\,)^{-1/2}\,\Vert\,\Big)\Vert\Psi\Vert.\label{t.3}
\end{eqnarray}
Thus $(\ref{t.2})$ holds.
Next we show that
\begin{eqnarray}
\Vert \,a_{R}(f)\,\Phi_{m}\,\Vert_{+1}^{2}\leq C_{m}\,\Vert \,a_{R}(f)\,\Phi_{m} \,\Vert+ o(R^{0}). \label{x}
\end{eqnarray}
Here $C$ is a constant independent of $R$. We have for $\eta \in Q(\hat{q}_{m}^{V})$,
\begin{eqnarray}
\hat{q}_{m}^{V} (\,\eta,\,a_{R}(f)\Phi_{m})&=&\frac{1}{2}\sum_{\mu=1}^{3}\left(\,(\,p_{\mu}+\sqrt{\alpha}\,\hat{A}_{\mu}\,)\eta,\,(\,p_{\mu}+\sqrt{\alpha}\,\hat{A}_{\mu}\,) a_{R}(f)\,\Phi_{m}\,\right)\nonumber \\
&&\quad+\left(\, \hhfm  ^{1/2}\, \eta,\,\hhfm  ^{1/2}\,a_{R}(f)\,\Phi_{m} \right) +\left(\, a_{R}^{\dagger}(\overline{f}) \, V_{+}^{1/2}\,\eta,\, V_{+}^{1/2}\,\Phi_{m}\,\right)\nonumber \\
&&\quad-\left(\, a_{R}^{\dagger}(\overline{f}) \, V_{-}^{1/2}\,\eta,\, V_{-}^{1/2}\,\Phi_{m}\,\right). \label{s}
\end{eqnarray}
Let $\Phi_{m}^{j}\in \mathcal{S}$ be such that $\Phi_{m}^{j}\to\Phi_{m}$ in $Q(\hat{q}_{m}^{0})$ as $j\to\infty$.
Then the first term of the right hand side of $(\ref{s})$ is computed as
\begin{eqnarray}
\lefteqn{\frac{1}{2}\sum_{\mu=1}^{3}\left(\,(\,p_{\mu}+\sqrt{\alpha}\hat{A}_{\mu}\,)\,\eta,\,(\,p_{\mu}+\sqrt{\alpha}\hat{A}_{\mu}\,)\,a_{R}(f)\,\Phi_{m}\,\right)}\nonumber \\
&&=\lim_{j\to\infty}\,\sum_{\mu=1}^{3}\Big\{ \frac{1}{2}\left(\,(\,p_{\mu}+\sqrt{\alpha }\hat{A}_{\mu}\,)\,\eta,\,a_{R}(f)\,(p_{\mu}+\sqrt{\alpha}\,\hat{A}_{\mu}\,)\,\Phi_{m}^{j}\right)\nonumber \\
&&\qquad \qquad \quad- \frac{1}{2}\left(\,(\,p_{\mu}+\sqrt{\alpha}\hat{A}_{\mu}\,)\,\eta,\,[\,a_{R}(f),\,\sqrt{\alpha}\hat{A}_{\mu}\,]\,\Phi_{m}^{j}\,\right) \Big\} \nonumber \\
&&=\frac{1}{2} \sum_{\mu=1}^{3}\left(\,a^{\dagger}_{R}(\overline{f})\,(\,p_{\mu}+\sqrt{\alpha }\hat{A}_{\mu}\,)\,\eta,\,(p_{\mu}+\sqrt{\alpha}\,\hat{A}_{\mu}\,)\,\Phi_{m}\right)\nonumber \\
&&\quad-\frac{\sqrt{\alpha}}{2}\sum_{\mu=1}^{3} \left((p_{\mu}+\sqrt{\alpha}\hat{A}_{\mu})\eta, R(\hhfm  +R)^{-1}(\overline{f},\overline{G}_{\mu}) \Phi_{m}\right)+E_{1}(R),\label{s.1}
\end{eqnarray}
where
\begin{eqnarray}
E_{1}(R)=-\!\lim_{j\to\infty}\! \frac{\sqrt{\alpha}}{2}\sum_{\mu=1}^{3}\left((p_{\mu}+\sqrt{\alpha}\hat{A}_{\mu})\eta,\, R[(\hhfm  +R)^{-1},\hat{A}_{\mu}]a(f)\Phi_{m}^{j}\,\right).
\end{eqnarray}
The first term of the right hand side of (\ref{s.1}) is computed as
\begin{eqnarray*}
\lefteqn{\frac{1}{2} \sum_{\mu=1}^{3}\left(\,a^{\dagger}_{R}(\overline{f})\,(\,p_{\mu}+\sqrt{\alpha }\hat{A}_{\mu}\,)\,\eta,\,(p_{\mu}+\sqrt{\alpha}\,\hat{A}_{\mu}\,)\,\Phi_{m}\right)}\nonumber \\
&&=\frac{1}{2} \sum_{\mu=1}^{3}\left(\,(\,p_{\mu}+\sqrt{\alpha }\hat{A}_{\mu}\,)\,a^{\dagger}_{R}(\overline{f})\,\eta,\,(p_{\mu}+\sqrt{\alpha}\,\hat{A}_{\mu}\,)\,\Phi_{m}\right)\nonumber \\
&&\quad+\frac{\sqrt{\alpha}}{2} \sum_{\mu=1}^{3}\left(\,[a^{\dagger}_{R}(\overline{f}),\,\hat{A}_{\mu}]\,\eta,\,(p_{\mu}+\sqrt{\alpha}\,\hat{A}_{\mu}\,)\,\Phi_{m}\right)\nonumber \\
\end{eqnarray*}
\begin{eqnarray}
\lefteqn{=\frac{1}{2} \sum_{\mu=1}^{3}\left(\,(\,p_{\mu}+\sqrt{\alpha }\hat{A}_{\mu}\,)\,a^{\dagger}_{R}(\overline{f})\,\eta,\,(p_{\mu}+\sqrt{\alpha}\,\hat{A}_{\mu}\,)\,\Phi_{m}\right)}\nonumber \\
&&\quad-\frac{\sqrt{\alpha}}{2} \sum_{\mu=1}^{3}\left(\,(\overline{G}_{\mu},\,\overline{f})\,R(\hhfm  +R)^{-1}\,\eta,\,(p_{\mu}+\sqrt{\alpha}\,\hat{A}_{\mu}\,)\,\Phi_{m}\right)+E_{2}(R),
\end{eqnarray}
where
\begin{eqnarray}
E_{2}(R)=\!-\!\lim_{j\to\infty}\frac{\sqrt{\alpha}}{2}\sum_{\mu=1}^{3}\left([\hat{A}_{\mu},R(\hhfm  +R)^{-1}]\eta,a(f)(p_{\mu}+\sqrt{\alpha}\hat{A}_{\mu})\Phi^{j}_{m}\right).
\end{eqnarray}
By (b) of Lemma \ref{Psi}, we see that $(f,\,G(\cdot,x))$ has a weak derivative with respect to $x$.
Therefore $(f,\,G_{\mu})\Phi_{m}\in D(|p|)$ and
\begin{eqnarray}
p \cdot (f,\,G)\Phi_{m} = (f,-i \nabla_{x} \cdot G) \Phi_{m}+(f,\,G)\cdot p \,\Phi_{m}. \label{s.7}
\end{eqnarray}
By (\ref{s.7}), we see that
\begin{eqnarray}
\lefteqn{\frac{1}{2}\sum_{\mu=1}^{3}\left(\,(\,p_{\mu}+\sqrt{\alpha}\hat{A}_{\mu}\,)\,\eta,\,(\,p_{\mu}+\sqrt{\alpha}\hat{A}_{\mu}\,)\,a_{R}(f)\,\Phi_{m}\,\right)}\nonumber \\
&&=\frac{1}{2} \sum_{\mu=1}^{3}\left(\,(\,p_{\mu}+\sqrt{\alpha }\hat{A}_{\mu}\,)\,a^{\dagger}_{R}(\overline{f})\,\eta,\,(p_{\mu}+\sqrt{\alpha}\,\hat{A}_{\mu}\,)\,\Phi_{m}\right) \nonumber \\
&&\quad -\sqrt{\alpha} \left(\,\eta,\, R\,(\,\hhfm  +R\,)^{-1}(\overline{f},\overline{G})\cdot (\,p+\sqrt{\alpha}\hat{A}\,) \,\Phi_{m}\right) \nonumber \\
&&\quad+\frac{i\sqrt{\alpha}}{2}\!\left(\eta, R(\hhfm  +R)^{-1}(\overline{f},\nabla_{x}\!\cdot\!\overline{G}) \Phi_{m}\right)+E_{1}(R)+E_{2}(R)+E_{3}(R),
\label{s.2}
\end{eqnarray}
where
\begin{eqnarray}
E_{3}(R)=-\frac{ \sqrt{\alpha}}{2}\sum_{\mu=1}^{3}\left(\,R\,[(\,\hhfm   +R\,)^{-1},\,\hat{A}_{\mu}\,] \eta,\,  (\overline{f},\overline{G_{\mu}} ) \,\Phi_{m}\right).
\end{eqnarray}
Let us show that
$
\lim_{R\to\infty}E_{i}(R)=0$
for $i=1,2,3$.
Note that
\begin{eqnarray*}
[\,\hat{A}_{\mu},\,(\,\hhfm   +R\,)^{-1}\,]
&=&(\,\hhfm   +R\,)^{-1}a(\omega_{m}G_{\mu})(\,\hhfm   +R\,)^{-1}\nonumber \\
&&\quad-(\,\hhfm   +R\,)^{-1}a^{\dagger}(\omega_{m}G_{\mu})(\,\hhfm   +R\,)^{-1}.
\end{eqnarray*}
Then
\begin{eqnarray}
{R\Vert\,[\,\hat{A}_{\mu},\,(\,\hhfm   +R\,)^{-1}\,]\eta\,\Vert}
&\leq& \Vert\,R \,(\,\hhfm   +R\,)^{-1}\Vert\,(\;\Vert\,a(\omega_{m}G_{\mu})\,(\,\hhfm   +R\,)^{-1}\,\eta\,\Vert \nonumber \\
&&\quad\quad+ \Vert\,a^{\dagger}(\omega_{m}G_{\mu})\,(\,\hhfm   +R\,)^{-1}\,\eta\,\Vert\;) \nonumber \\
&\leq& 2\Vert\,\omega_{m}^{1/2} \,G_{\mu}\,\Vert\,\Vert\,(1+\hhfm  )^{1/2} (\,\hhfm   +R\,)^{-1}\,\eta\,\Vert\nonumber \\
&& \rightarrow 0\;\text{ as }R\to\infty.
\end{eqnarray}
Thus
\begin{eqnarray}
\lim_{R\to\infty}E_{3}(R)=0.
\end{eqnarray}
Also $E_{1}(R)$ can be estimated as
\begin{eqnarray}
\lefteqn{|E_{1}(R)|\leq\frac{\sqrt{\alpha}}{2} \sum_{\mu=1}^{3}\left\Vert\,R(\hhfm  +R)^{-1}(p_{\mu}+\sqrt{\alpha}\hat{A}_{\mu})\eta\,\right\Vert\times}\nonumber\\
&&\quad\,\limsup_{j\to\infty}\left\Vert\,(a(\omega_{m}^{1/2}G_\mu)+a^{\dagger}(\omega_{m}^{1/2}G_\mu))(\hhfm  +R)^{-1}a(f)\Phi_{m}^{j}\,\right\Vert.
\end{eqnarray}
Since
\begin{eqnarray}
\lefteqn{\left\Vert\,(a(\omega_{m}^{1/2}G_\mu)+a^{\dagger}(\omega_{m}^{1/2}G_\mu))(\hhfm  +R)^{-1}a(f)\Phi_{m}^{j}\,\right\Vert}\nonumber \\
&&\leq \textrm{const.}\Vert\,(\hhfm  +1)^{1/2}(\hhfm  +R)^{-1}\,\Vert\lim_{j\to\infty}\Vert \,a(f)\Phi_{m}^{j}\,\Vert
\end{eqnarray}
and
\begin{eqnarray}
\lim_{R\to\infty}\,\Vert\,(\hhfm  +1)^{1/2}(\hhfm  +R)^{-1}\,\Vert=0,
\end{eqnarray}
it follows that
\begin{eqnarray}
\lim_{R\to\infty}E_{1}(R)=0.
\end{eqnarray}
Similarly $E_{2}(R)$ can be estimated as
\begin{align}
|E_{2}(R)| &\leq \textrm{const.}\Vert\,R(\hhfm  +1)^{1/2}(\hhfm  +R)^{-1}\eta\Vert\times \nonumber \\
&\qquad \limsup_{j\to\infty} \Vert (\hhfm  +R)^{-1}a(f)\,(p_{\mu}+\sqrt{\alpha}\hat{A}_{\mu})\Phi_{m}^{j}\Vert\nonumber \\
&\leq \textrm{const.}\Vert\,(\hhfm  +1)^{1/2}\eta\Vert\,\Vert a^{\dagger}(\overline{f})\,(\hhfm  +R)^{-1}\Vert\,\Vert(p_{\mu}+\sqrt{\alpha}\hat{A}_{\mu})\Phi_{m}\Vert\nonumber \\
&\leq \textrm{const.}\Vert\,(\hhfm  +1)^{1/2} \eta\Vert \, \Vert\,(\hhfm  +1)^{1/2}(\hhfm  +R)^{-1}\Vert\,
\Vert(p_{\mu}+\sqrt{\alpha}\hat{A}_{\mu})\Phi_{m}\Vert.
\end{align}
Thus
\begin{eqnarray}
\lim_{R\to\infty}E_{2}(R)=0.
\end{eqnarray}
Therefore by (\ref{s.2}),
\begin{eqnarray}
\lefteqn{\frac{1}{2}\sum_{\mu=1}^{3}\left(\,(\,p_{\mu}+\sqrt{\alpha}\hat{A}_{\mu}\,)\,\eta,\,(\,p_{\mu}+\sqrt{\alpha}\hat{A}_{\mu}\,)\,a_{R}(f)\,\Phi_{m}\,\right)}\nonumber \\
=&&\frac{1}{2}\, \sum_{\mu=1}^{3}\left(\,(\,p_{\mu}+\sqrt{\alpha }\hat{A}_{\mu}\,)\,a^{\dagger}_{R}(\overline{f})\,\eta,\,(p_{\mu}+\sqrt{\alpha}\,\hat{A}_{\mu}\,)\,\Phi_{m}\right) \nonumber \\
&&\quad-\sqrt{\alpha} \left(\,\eta,\, R\,(\,\hhfm  +R\,)^{-1}(\overline{f},\overline{G})\cdot(\,p+\sqrt{\alpha}\hat{A}\,) \,\Phi_{m}\right) \nonumber \\
&&\quad+\frac{i\sqrt{\alpha}}{2}\left(\,\eta,\, R\,(\,\hhfm  +R\,)^{-1}(\overline{f},\nabla_{x}\cdot\overline{G}) \,\Phi_{m}\right) +o(R^{0}).\label{s.3}
\end{eqnarray}
Next let us see the second term of the right hand side of $(\ref{s})$.
\begin{eqnarray}
\lefteqn{(\,\hhfm  ^{1/2}\,\eta,\,\hhfm  ^{1/2}\,a_{R}(f)\,\Phi_{m})}\nonumber \\
&&=\lim_{j\to\infty}\,(\,\eta,\,a_{R}(f)\,\hhfm  \,\Phi_{m}^{j}\,) - \lim_{j\to\infty}(\,\eta,\, R\,(\,\hhfm  +R\,)^{-1} \,a(\omega_{m}f) \,\Phi_{m}^{j}\,)\nonumber \\
&&=(\,\hhfm  ^{1/2}\,a_{R}^{\dagger}(\overline{f})\,\eta,\,\hhfm  ^{1/2}\, \Phi_{m}\,)- (\,\eta,\, R\,(\,\hhfm  +R\,)^{-1} \,a(\omega_{m}f) \,\Phi_{m}\,). \nonumber \\ \label{s.4}
\end{eqnarray}
Therefore by (\ref{s}), (\ref{s.3}) and (\ref{s.4}), we see that
\begin{eqnarray}
\lefteqn{\hat{q}_{m}^{V} (\,\eta,\,a_{R}(f)\Phi_{m})-E^{V}_{m}(\,\eta,\,a_{R}(f)\Phi_{m})}\nonumber \\
&&=\hat{q}_{m}^{V} (\,a_{R}^{\dagger}(\overline{f})\,\eta,\,\Phi_{m})-E^{V}_{m}(\,a_{R}^{\dagger}(\overline{f})\,\eta,\,\Phi_{m})\nonumber \\
&&\quad-\sqrt{\alpha} \left(\,\eta,\, R\,(\,\hhfm  +R\,)^{-1}(\overline{f},\overline{G})\cdot(\,p+\sqrt{\alpha}\hat{A}\,) \,\Phi_{m}\right) \nonumber \\
&&\quad+\frac{i\sqrt{\alpha}}{2}\left(\,\eta,\, R\,(\,\hhfm  +R\,)^{-1}(\overline{f},\nabla_{x}\cdot\overline{G}) \,\Phi_{m}\right) \nonumber \\
&&\quad-(\,\eta,\,R\,(\,\hhfm  +R\,)^{-1} \,a(\omega_{m}f) \,\Phi_{m}\,)+o(R^{0}).\label{s.5}
\end{eqnarray}
Since $\Phi_{m}$  is the eigenvector of $\hat{H}_{m}^{V}$, the first line of the right hand side of (\ref{s.5}) is vanished.
Now take $\eta = a_{R}(f)\,\Phi_{m}$ in (\ref{s.5}). Since $\Vert\,R\,(\,\hhfm  +R\,)^{-1}\Vert\leq 1 $, by (\ref{s.5}), we obtain that
\begin{eqnarray}
\lefteqn{\Vert\,a_{R}(f)\,\Phi_{m} \,\Vert_{+1} \leq \Vert a_{R}(f)\,\Phi_{m} \,\Vert\,
\Big\{\,\sqrt{\alpha} \Vert \, (\overline{f},\overline{G})\cdot (\,p+\sqrt{\alpha}\hat{A}\,) \,\Phi_{m}\,\Vert }\nonumber \\
&&\quad+\frac{\sqrt{\alpha}}{2}\Vert \, (\overline{f},\nabla_{x}\cdot\overline{G}\,) \,\Phi_{m}\,\Vert +\Vert\,a(\omega_{m}f) \,\Phi_{m} \,\Vert + 1 \, \Big\}+o(R^{0}).\label{s.6}
\end{eqnarray}
Thus $(\ref{x})$  is proven. It is seen that as $R\to \infty$,   $a_{R}(f)\Phi_{m}\to a(f)\Phi_{m}$.
Therefore by (\ref{s.6}), we see that $a(f)\Phi_{m}\in Q(\hat{q}_{m}^{V})$, since $Q(\hat{q}_{m}^{V})$ is complete.
Letting $R\to\infty$ in (\ref{s.5}), we complete the lemma.\qed
\end{pf}
\begin{lem}\label{number b'dd}
For any $\theta=(\theta_{1},\,\theta_{2})\in L^{\infty}(\mathbb{R}^{3};\mathbb{R}^{2})$,
\begin{eqnarray}
\Vert \,d\Gamma(\,\theta^{2}\,)^{1/2}\,\Phi_{m}\,\Vert^{2} \leq C\alpha \,\sum_{\mu,j}\,\int_{\mathbb{R}^{3}}\, \frac{|\hat{\varphi}_{j}^{\mu}(k)|^{2}\,\theta_{j}(k)^{2}}{\omega(k)\,\omega_{m}(k)^{2}}\,dk,
\end{eqnarray}
where $C$ is a constant independent of $\alpha$ and $m \in (0,m_0)$.
In particular,
\begin{align}
\sup_{0<m<m_0} \Vert N^{1/2}\Phi_m\Vert<\infty.
\end{align}
\end{lem}
\begin{pf}
Inserting $\eta=a(f)\Phi_{m}$ into (\ref{s.8}), we have
\begin{eqnarray}
(\,a(f)\,\Phi_{m},\,a(\omega_{m}f)\,\Phi_{m}\,)&\leq&\lefteqn{\, -\sqrt{\alpha}\left( \,a(f)\,\Phi_{m},\,(\overline{f},\overline{G})\cdot (\,p+\sqrt{\alpha}\hat{A}\,)\,\Phi_{m} \right)} \nonumber \\
&&\quad +\frac{i\sqrt{\alpha}}{2}\left( a(f)\,\Phi_{m},(\overline{f},\,\nabla\cdot\overline{G}\,)\,\Phi_{m}\,\right). \label{p.t.}
\end{eqnarray}
Let $\{g_{i}\}$ be a complete orthonormal system of $L^{2}(\mathbb{R}^{3};\mathbb{C}^{2})$ such that each $g_{i}\in D(\omega_{m}^{1/2})$.
Let $f=\omega_{m}^{-1/2}\theta g_{i}$.
The following identity follows and its proof will be given in Appendix C:
\begin{eqnarray}
\sum_{i}\left(\,a(\omega_{m}^{-1/2}\,\theta\,g_{i})\,\Phi_{m},\,a(\omega_{m}^{1/2}\,\theta\,g_{i})\,\Phi_{m}\, \right)&=&\sum_{j=1,2}\int_{\mathbb{R}^{3}}\theta_{j}(k)^{2}\Vert\,(a_{j}\Phi_{m})(k)\,\Vert^{2}\,dk\nonumber \\
&=&\left\Vert\, d\Gamma(\,\theta^{2}\,)^{1/2}\,\Phi_{m}\, \right\Vert^{2}. \label{y}
\end{eqnarray}
By  $(\ref{p.t.})$  and  $(\ref{y})$,  we have
\begin{eqnarray}
\lefteqn{\Vert \,d\Gamma (\,\theta^{2}\,)^{1/2} \,\Phi_{m} \,\Vert^{2}} \nonumber \\
&&\leq -\sqrt{\alpha}\sum_{i,\mu}\, \left(\,a(\,\omega_{m}^{-1/2}\,\theta\,g_{i}\,)\,\Phi_{m},\,(\,\omega_{m}^{-1/2}\,\theta\,\overline{g_{i}},\,\overline{G}_{\mu}\,)\,(\,p_{\mu}+\sqrt{\alpha}\hat{A}_{\mu}\,)\,\Phi_{m}\,\right) \nonumber \\
&&\quad+\frac{i\sqrt{\alpha}}{2}\,\left(\,a(\,\omega_{m}^{-1/2}\,\theta\,g_{i}\,)\,\Phi_{m},\,(\,\omega_{m}^{-1/2}\,\theta\,\overline{g_{i}},\,\nabla_{x}\cdot\overline{G}_{\mu}\,)\,\Phi_{m}\,\right) \nonumber \\
&&=-\sqrt{\alpha}\sum_{\mu}\left(\,a(\,\omega_{m}^{-1}\,\theta^{2}\, \overline{G_{\mu}}\,)\,\Phi_{m},\,(\,p_{\mu}+\sqrt{\alpha}\hat{A}_{\mu}\,)\,\Phi_{m}\,\right)\nonumber \\
&&\quad+ \frac{i\sqrt{\alpha}}{2}\left(\,a(\,\omega_{m}^{-1}\,\theta^{2}\,\nabla_{x}\cdot \overline{G}\,)\,\Phi_{m},\,\Phi_{m}\, \right) \nonumber\\
&&=\int_{\mathbb{R}^{3}}\,\Big\{\,-\sqrt{\alpha}\sum_{\mu} \left(\,\omega_{m}(k)^{-1}\,\theta (k)^{2}\,\overline{G_{\mu}(k)}\,a(k)\,\Phi_{m},\,(\,p_{\mu}+\sqrt{\alpha}\hat{A}_{\mu}\,)\,\Phi_{m}\,\right)_{\mathcal{H}}\nonumber \\
&&\quad\quad+\frac{i\sqrt{\alpha}}{2}\left(\,\omega_{m}^{-1}(k)\,\theta (k)^{2}\,\nabla_{x}\cdot \overline{G (k)}\,a(k)\,\Phi_{m},  \,\Phi_{m}\,\right)_{\mathcal{H}} \,\Big\}\,dk.
\end{eqnarray}
This calculation can be checked in a similar way to the proof of (\ref{y}). Thus by the Schwartz inequality and the inequality of arithmetic and geometric means,
we see that
\begin{eqnarray}
\Vert \,d\Gamma (\theta^{2})^{1/2} \,\Phi_{m} \,\Vert^{2}
&\leq&
\frac{1}{2}\sum_{j=1,2}\int_{\mathbb{R}^{3}} \,\theta_{j}(k)^{2}\Vert\, (a_{j}\Phi_{m})(k) \,\Vert^{2}dk \nonumber \\
&&\quad +\alpha\,\int_{\mathbb{R}^{3}}\, \omega_{m}(k)^{-2}\,\Vert\,\theta(k) G(k)\cdot (\,p+\sqrt{\alpha}\hat{A}\,)\,\Phi_{m}\,\Vert^{2}\,dk \nonumber \\
&&\quad\quad + \frac{\alpha}{4}\int_{\mathbb{R}^{3}} \,\omega_{m}(k)^{-2} \,\Vert\, \theta (k)\, \nabla_{x} \cdot G(k)\, \Phi_{m}\, \Vert^{2}\, dk.
\end{eqnarray}
Thus we have
\begin{eqnarray}
 \Vert \, d\Gamma (\,\theta^{2}\,)^{1/2}\, \Phi_{m} \,\Vert^{2}
 &\leq&
   2\alpha\,\int_{\mathbb{R}^{3}} \,\omega_{m}(k)^{-2}\,\Vert \,\theta(k) G(k)\cdot (p+\sqrt{\alpha}\hat{A})\,\Phi_{m}\,\Vert^{2}\,dk \nonumber \\
&&\qquad + \frac{\alpha}{2} \int_{\mathbb{R}^{3}} \,\omega_{m}(k)^{-2} \,\Vert\,\theta(k) \nabla_{x}\cdot G(k)\, \Phi_{m} \,\Vert^{2}\, dk\nonumber .
\end{eqnarray}
Since by (a) and (b) of Lemma \ref{Psi}, $\Psi (k,x)$ and $\hat{\varphi}(k)\partial_{x,\mu} \Psi(k,x)$ are bounded, we have
\begin{eqnarray}
&&\Vert\, \theta(k)\,G(k)\cdot (p+\sqrt{\alpha}\hat{A})\,\Phi_{m}\,\Vert^{2}\leq \mathrm{const.}\,\sum_{\mu,j}\,\frac{|\hat{\varphi}_{j}^{\mu}(k)|^{2}\theta_{j}(k)}{\omega(k)}\nonumber, \\
&&\Vert\, \theta(k)\,\nabla_{x}\cdot G(k)\,\Phi_{m} \,\Vert^{2}\leq \mathrm{const.}\,\sum_{\mu,j}\,\frac{|\hat{\varphi}_{j}^{\mu}(k)|^{2}\theta_{j}(k)}{\omega(k)}.
\end{eqnarray}
Thus the lemma follows. \qed
\end{pf}
\subsection{Derivative of massive ground states}
\begin{lem} \label{derivative b'dd}
Suppose Assumption \ref{as.exp}. Let $1\leq p<2$. Then
for all $n\geq 0$, $\Phi_{m}^{(n)}\in H^{1}(\mathbb{R}^{3+3n})$ and $\{\Vert \Phi_{m}^{(n)}\Vert_{W^{1,p}(\Omega)}\}_{m}$ is bounded for sufficiently small $m>0$
and for any measurable, bounded set $\Omega\subset \mathbb{R}^{3+3n}$.
Here $H^{1}(\mathbb{R}^{3+3n})$ and $W^{1,p}(\Omega)$ are the Sobolev spaces.
\end{lem}
\begin{pf}
For a function $f:\mathbb{R}^{3}\to\mathbb{C}^{2}$, we define the shift $t_{h}$ and the difference $\delta_{h}$ by
\begin{eqnarray}
t_{h}f(\cdot)=f(\cdot+h),\quad \delta_{h}f=t_{h}f-f.
\end{eqnarray}
We define furthermore  bounded operators $T_{h}$ and $\Delta_{h}$ on $L^{2}(\mathbb{R}^{3};\mathbb{C}^{2})$ by
\begin{eqnarray}
T_{h}\psi(\cdot)=\psi(\cdot+h),\quad \Delta_{h}\psi=T_{h}\psi-\psi,\quad \psi\in L^{2}(\mathbb{R}^{3};\mathbb{C}^{2}).
\end{eqnarray}
Also we define
\begin{eqnarray}
\tau_{h}a\Psi(k)=a\Psi(k+h), \quad \mathfrak{d}_{h}a\Psi(k)=a\Psi(k+h)-a\Psi(k).
\end{eqnarray}
We see that $T_{h}^{*}=T_{-h}$ and $\Delta_{h}^{*}=\Delta_{-h}$. Let $f$ be such that $\delta_{h}f\in D(\omega_{m})$ for all $h\in \mathbb{R}^{3}$.
By (\ref{p.t.}) with $f$ replaced  by $\delta_{-h}f$
and by the identity
$\delta_{-h}(\omega_{m} f)=\delta_{-h}\omega_{m}\cdot t_{-h}f+\omega_{m}\delta_{-h}f$,
we have
\begin{align}
&\left(\, a(\delta_{-h}f)\,\Phi_{m},\,a(\delta_{-h}(\omega_{m} f)\,)\,\Phi_{m}\,\right)\nonumber \\
& \leq \left(\, a(\delta_{-h}f)\Phi_{m},a(\delta_{-h}\omega_{m}\cdot t_{-h}f)\,\Phi_{m})\,\right)\nonumber \\
& -\sqrt{\alpha}\left(a(\delta_{-h}f)\Phi_{m},(f,\,\delta_{h}G)\cdot (p+\sqrt{\alpha}\hat{A})\Phi_{m}\right)+\frac{i\sqrt{\alpha}}{2}\left(a(\delta_{-h}f)\Phi_{m},(f,\delta_{h}\nabla_{x}\cdot G) \Phi_{m}\right),
\end{align}
where
$$\delta_{h}G(k)=G(k+h,x)-G(k,x)$$
and
$$\delta_{h}\nabla_{x}\cdot G(k)=(\nabla_{x}\cdot G)(k+h,x)-(\nabla_{x}\cdot G)(k,x).$$
Let $f=\omega_{m}^{-1/2}\theta g_{i}$ and suppose that $\delta_{h}(f)$ and $f$ is in $D(\omega_{m})$.
Note that
\[
\Vert\,d\Gamma(\,\Delta_{h}^{*}\theta^{2}\Delta_{h})^{1/2}\Phi_{m}\,\Vert^2=\int_{\mathbb{R}^{3}}\Vert \,\theta(k)^{2}\mathfrak{d}_{h}a\Phi_{m}(k)\,\Vert^{2}\,dk.
\]
Then in a similar way to the proof of Lemma $\ref{number b'dd}$, we obtain that
\begin{eqnarray}
\lefteqn{\Vert \,d\Gamma (\Delta_{h}^{*}\,\theta^{2}\,\Delta_{h})^{1/2}\,\Phi_{m} \, \Vert^{2}} \nonumber \\
&&\leq 2\sum_{i}\,\left(\, a(\delta_{-h}(\omega_{m}^{-1/2}\theta g_{i}))\Phi_{m},\,a(\delta_{-h}\omega_{m}\cdot t_{-h}(\omega_{m}^{-1/2}\theta g_{i}))\,\Phi_{m})\,\right) \nonumber \\
&&\quad + 2\alpha\,\int_{\mathbb{R}^{3}} \,\omega_{m}(k)^{-2} \left\Vert \theta(k) \delta_{h}G(k)\cdot ( p+\sqrt{\alpha}\hat{A})\Phi_{m}\right\Vert^{2}dk \nonumber \\
&&\quad + \frac{\alpha}{2}\int_{\mathbb{R}^{3}} \,\omega_{m}(k)^{-2} \,\left\Vert \theta(k) \delta_{h}\nabla_{x}\cdot G(k) \,\Phi_{m} \right\Vert^{2} \,dk. \label{rr}
\end{eqnarray}
The first term of the right hand side of $(\ref{rr})$ is rewritten as
\begin{eqnarray}
2\int_{\mathbb{R}^{3}} \omega_{m}(k)^{-1}t_{h}\delta_{-h}\omega_{m}(k)\left(\theta(k)\mathfrak{d}_{h}a\Phi_{m}(k),\,\theta(k)\tau_{h}a\Phi_{m}(k)\right)dk.\label{rr.1}
\end{eqnarray}
This can be proven in a similar way to that of $(\ref{y})$.
By Schwartz's inequality and the inequality of arithmetic and geometric means, (\ref{rr.1}) can be estimated from above as
\begin{eqnarray}
\lefteqn{|\text{(\ref{rr.1})}|\leq\frac{1}{2}\left\Vert \, d\Gamma \left(\, \Delta_{h}^{*}\,\theta^{2}\,\Delta_{h}\,\right)^{1/2}\,\Phi_{m} \,\right\Vert^{2} }\nonumber \\
&&+2\,\int_{\mathbb{R}^{3}}\,\omega_{m}(k)^{-2}t_{h}\delta_{-h}\omega_{m}(k)^{2} \,\Vert\,\theta (k)\,\tau_{h}a\Phi_{m}(k)\,\Vert^{2}\,dk.\;\label{rr.3}
\end{eqnarray}
We also see that
\begin{eqnarray}
\lefteqn{\Vert \,\delta_{h}G(k)\cdot(\,p+\sqrt{\alpha}\hat{A}\,)\,\Phi_{m}\,\Vert^{2}} \nonumber \\
&&\leq \Vert \,\delta_{h} \nabla_{x} \cdot G(k) \,\Phi_{m}\,\Vert^{2} + C_{0}\;\sum_{\mu=1}^{3} \Vert\,\delta_{h}G_{\mu}(k)\,\Phi_{m}\,\Vert_{+1}^{2} \nonumber \\
&&= \Vert \,\delta_{h}\nabla_{x} \cdot G(k) \,\Phi_{m}\,\Vert^{2} 
+C_{0}\,\sum_{\mu=1}^{3}\,\left( \frac{1}{2}\Vert \,|\; \delta_{h} \,\nabla_{x} G_{\mu}(k)\;| \,\Phi_{m} \,\Vert^{2}+ \Vert\,\delta_{h}G_{\mu}(k)\,\Phi_{m}\,\Vert^{2}\, \right).\nonumber\\
 \label{rr.4}
\end{eqnarray}
Here $C_{0}$ is a constant independent of $k$ and sufficiently small $m$.
Therefore by (\ref{rr}), (\ref{rr.3}) and (\ref{rr.4}), we see that
\begin{eqnarray}
\lefteqn{\Vert \, d\Gamma (\,\Delta^{*}_{h}\theta^{2}\Delta_{h}\,)^{1/2}\,\Phi_{m} \, \Vert^{2}} \nonumber \\
&&\leq C_{1}\int_{\mathbb{R}^{3}}dk\;\omega_{m}(k)^{-2}
\Big\{\, \sum_{\mu=1}^{3}\big(\;\Vert\,\theta(k)\, \delta_{h}G_{\mu}(k)\Phi_{m}\,\Vert^{2}
  +  \Vert \,\theta(k)\, |\,\delta_{h}\nabla G_{\mu}(k)\,| \,\Phi_{m}\, \Vert^{2}\;\big) \nonumber \\
&&\quad  +  \!\Vert  \theta(k)\delta_{h}\nabla_{x}\cdot G(k) \Phi_{m} \Vert^{2}
\!+ \!\Vert\theta (k)\tau_{h}a(k)\Phi_{m}\Vert^{2}|\omega_{m}(k+h)-\omega_{m}(k)|^{2} \Big\}. \label{rr.2}
\end{eqnarray}
Here $C_{1}$ is a constant independent of $k$ and sufficiently small $m$ but depends on $\alpha$.
Since $\theta$ is an arbitrary bounded, $\mathbb{R}^{2}$-valued function, (\ref{rr.2}) implies that there exists a null set $N_{0}\subset \mathbb{R}^{3}\times\mathbb{R}^{3}$ so that if $(k,h)\notin N_{0}$, it holds that
\begin{eqnarray}
\Vert \, \mathfrak{d}_{h}a\Phi_{m}(k) \,\Vert^{2}
&\leq&  \frac{C_{1}}{\omega_{m}(k)^{2}} \Big(\,\sum_{\mu=1}^{3}(\;\Vert \delta_{h}G_{\mu}(k)\Phi_{m}\,\Vert^{2}
  +  \Vert \, |\,\delta_{h}\nabla_{x} G_{\mu}(k)\,| \,\Phi_{m}\, \Vert^{2} \;)\nonumber \\
&&  +  \Vert \,\delta_{h}\nabla_{x}\cdot G(k) \Phi_{m} \Vert^{2} \!+\!  \Vert \tau_{h}a(k)\Phi_{m}\Vert^{2} \left| \omega_{m}(k+h)-\omega_{m}(k) \right|^{2}\Big).\nonumber\\
\label{null}\end{eqnarray}
Thus there exists a null set $N_{1}\subset\mathbb{R}^{3}$  such that
if $k\notin N_{1}$, there exists a null set $N_{2}(k)$ and (\ref{null}) holds whenever $h\notin N_{2}(k)$.
Moreover by Lemma \ref{number b'dd}, we can see that there exists a null set $N_{3}$ so that if $k+h\notin N_{3}$, it holds that
\begin{eqnarray}
\Vert\,\tau_{h}a\Phi_{m}(k)\,\Vert^{2}\leq \sum_{j,\mu}\frac{|\hat{\varphi}_{j}^{\mu}(k+h)|^{2}}{\omega_{m}(k+h)^{2}\omega(k+h)}.
\end{eqnarray}
Therefore we obtain that for $k\notin N_{1}$, $k+h\notin N_{3}$ and $h\neq 0$,
\begin{eqnarray}
\lefteqn{\Vert\, |h|^{-1}(a\Phi_m(k+h)-a\Phi_{m}(k)\, \Vert^{2}}\nonumber \\
&&\leq  \frac{C_{1}}{\omega_{m}(k)^{2}} \Big(\,\sum_{\mu=1}^{3}(\; \Vert\, |h|^{-1}\delta_{h}G_{\mu}(k)\Phi_{m}\,\Vert^{2}
  +  \Vert \, |\,|h|^{-1}\delta_{h}\nabla_{x} G_{\mu}(k)\,| \,\Phi_{m}\, \Vert^{2}\;) \nonumber \\
&&\quad +  \Vert|h|^{-1}\delta_{h}\nabla_{x}\cdot G(k) \Phi_{m} \Vert^{2} + \sum_{j,\mu}\frac{ |\hat{\varphi}_{j}^{\mu}(k+h)|^{2} }{\omega_{m}(k+h)^{2}\omega(k+h) } \frac{|\omega(k+h)-\omega(k)|^{2}}{|h|^{2}}\Big).\qquad \label{rr.5}
\end{eqnarray}
By (c) of Lemma \ref{Psi}, Theorem \ref{exp} and $(\ref{rr.5})$, we see that
for $0\neq k\notin N_{1}$, there exist positive constants $h_{0}(k)$ and $C$ independent of $k$ and sufficiently small $m$ so that
\begin{eqnarray}
\lefteqn{\Vert \,|h|^{-1}(a\Phi_m(k+h)-a\Phi_m(k))\, \Vert^{2}}\nonumber \\
&&\leq C \omega_{m}(k)^{-2}\sum_{\nu,j}\left(\;(1+|k|^{-3})|\hat{\varphi}_{j}^{\nu}(k)|^{2}+\sum_{\lambda}|k|^{-1}\,|\,\partial_{\lambda}\hat{\varphi}_{j}^{\nu}(k)\,|^{2}\,\right).
\label{aaa}
\end{eqnarray}
holds if $0<|h|<h_{0}(k)$ and $h\notin N_{2}(k)\cup (N_{3}- k)$. Here $N_{3}-k=\{h| h+k\in N_{3}\}$.
Let $e_{\mu}$, $\mu=1,2,3$ be the canonical basis: $e_{1}=\!{}^T\!(1,0,0)$, $e_{2}=\!{}^T\!(0,1,0)$, $e_{3}=\!{}^T\!(0,0,1)$.
By (\ref{aaa}) and the Alaoglu theorem, for $\mathrm{a.e.}$ $k$, there exists a sequence $\{h_{l}\}_{l=1}^{\infty}$ so that $h_{l}\to 0$, $l\to\infty$ and $|h_{l}|^{-1}\,\mathfrak{d}_{-|h_{l}|e_{\mu}}a(k)\,\Phi_{m}$
is weakly converges in $\mathcal{H}$. Here the sequence $\{h_{l}\}$ depends on $k$. We set
\begin{eqnarray}
v_{\mu}(k)=\mathrm{w}\text{-}\,\lim_{l\to \infty}\,|\,h_{l}\,|^{-1}\,\mathfrak{d}_{-|h_{l}|e_{\mu}} a\Phi_{m}(k).\label{v(k)}
\end{eqnarray}
By $(\ref{aaa})$ again, we have
\begin{eqnarray}
\Vert v_{\mu}(k) \Vert^{2} \leq C \omega_{m}(k)^{-2}\sum_{\nu,j}\left(\,(1+|k|^{-3})|\hat{\varphi}_{j}^{\nu}(k)|^{2}+\sum_{\lambda}|k|^{-1}|\,\partial_{\lambda}\hat{\varphi}_{j}^{\nu}(k)\,|^{2}\right). \label{v_{mu}}
\end{eqnarray}
Next, we will show that for $n\geq 1$, $\Phi_{m}^{(n)}\in L^{2}(\mathbb{R}^{3+3n};\mathbb{C}^{2})$ is weakly differentiable and
\begin{eqnarray}
D_{k(i,\mu)}\,\Phi_{m}^{(n)} = \frac{1}{\sqrt{n}}\,v_{\mu}^{(n-1)}(k_{i})(x,k_{1},\ldots,\widehat{k_{i}},\ldots,k_{n}).
\end{eqnarray}
Here $D_{k(i,\mu)}$ denotes the weak derivative with respect to $k_{i,\mu}$, ($i=1,\ldots n$, $\mu=1,2,3$) and
$\widehat{k_{i}}$ denotes omitting $k_{i}$.
Since $\Phi_{m}^{(n)}(x,k_{1},\ldots,k_{n})$ is a symmetric function for variables $k_{1},\ldots,k_{n}$, it suffices to consider
the derivative in $k_{1,\mu}$.
Let $u\in C_{c}^{\infty}(\mathbb{R}^{3+3n};\mathbb{C}^{2})$.
We have by the definition of $a(k)$, (\ref{N1}), (\ref{N2}) and the dominated convergence theorem,
\begin{eqnarray}
\lefteqn{\int_{\mathbb{R}^{3+3n}}\, (\partial_{k_{1},\mu}u)(x,k_{1},\ldots ,k_{n}) \,\Phi_{m}^{(n)}(x,k_{1},\ldots,k_{n})\,dxdk_{1}\ldots dk_{n} }\nonumber \\
&&=\frac{1}{\sqrt{n}}\,\int_{\mathbb{R}^{3}} dk_{1}\, \lim_{|h|\to 0}\,\int_{\mathbb{R}^{3+3(n-1)}} dxdk_{2} \ldots dk_{n}\nonumber \\
&&\,\frac{|\,u(x,k_{1}+|h|e_{\mu},\ldots,k_{n})-u(x,k_{1},\ldots,k_{n})|}{|h|} (a\Phi_{m}(k_{1}))^{(n-1)}(x,k_{2},\ldots,k_{n}).\nonumber \\ \label{rr.6}
\end{eqnarray}
The right hand side of (\ref{rr.6}) can be computed as
\begin{eqnarray}
\lefteqn{\frac{1}{\sqrt{n}}\,\int_{\mathbb{R}^{3}}\, dk_{1} \lim_{|h|\to 0}\, \int_{\mathbb{R}^{3+3(n-1)}}dxdk_{2}\ldots dk_{n}}\nonumber \\
&&\quad u(x,k_{1},\ldots,k_{n})(|h|^{-1}\mathfrak{d}_{-|h|e_{\mu}}a\Phi_{m}(k_{1})\;)^{(n-1)}\,(x,k_{2},\ldots,k_{n})  \nonumber \\
&&=\frac{1}{\sqrt{n}}\,\int_{\mathbb{R}^{3}}\,dk_{1}
\lim_{l\to \infty} \!(u(\cdot,k_{1},\ldots),
|h_{l}|^{-1}(\mathfrak{d}_{-|h_{l}|e_{\mu}}a\Phi_{m}(k_{1}))^{(n-1)})_{L^{2}(\mathbb{R}^{3+3(n-1)})}.\ \ \
\end{eqnarray}
Thus by the definition of $v_{\mu}$, we have
\begin{eqnarray}
\lefteqn{\int_{\mathbb{R}^{3+3n}}\, (\partial_{k_{1},\mu}u)(x,k_{1},\ldots ,k_{n}) \,\Phi_{m}^{(n)}(x,k_{1},\ldots,k_{n})\,dxdk_{1}\ldots dk_{n} } \nonumber \\
&&\!\!\!=\!\!\!\int_{\mathbb{R}^{3+3n}} \!\!\!\!\!\!u(x,k_{1},\ldots ,k_{n})\!\frac{1}{\sqrt{n}}v_{\mu}^{(n-1)}(k_{1})(x,k_{2},\ldots,k_{n}) dxdk_{1}\ldots dk_{n}.
\end{eqnarray}
By the infrared regularity condition (\ref{I.F.}), we have
\begin{eqnarray}
\lefteqn{\int_{\mathbb{R}^{3+3n}}\, |\,v_{\mu}(k_{1})^{(n-1)}(x,k_{2},\ldots,k_{n})\,|^{2}\,dxdk_{1}\ldots dk_{n}}\nonumber \\
&&\leq C\int_{\mathbb{R}^{3}}\,\omega_{m}(k_{1})^{-2}\sum_{\nu,j}\left(\,(1+|k_{1}|^{-3})|\hat{\varphi}_{j}^{\nu}(k_{1})|^{2}+\sum_{\lambda}|k_{1}|^{-1}|\,\partial_{\lambda}\hat{\varphi}_{j}^{\nu}  (k_{1})\,|^{2}\right)dk_{1}\nonumber \\
&&<\infty.
\end{eqnarray}
Let $\Omega\subset\mathbb{R}^{3}$ be a measurable set such that $\Omega\subset\{x|\;|x|<R\}$ for some $R>0$.
By Assumption\! \ref{varphi}, Schwartz's inequality and (\ref{v_{mu}}), we have
\begin{eqnarray}
&&{\int_{\Omega}\, |\,v_{\mu}(k_{1})^{(n-1)}(x,k_{2},\ldots,k_{n})\,|^{p}\,dxdk_{1}\ldots dk_{n} }\nonumber \\
&&\leq C' \!\int_{\mathbb{R}^{3}}\!dk_{1}\!\! \left(\int_{|x|,|k_{2}|,\ldots,|k_{n}|<R} \!|v_{\mu}(k_{1})^{(n-1)}(x,k_{2},\ldots,k_{n})|^{2}dxdk_{2}\ldots dk_{n}\!\! \right)^{p/2}\nonumber \\
&&\leq C' \!\int_{\mathbb{R}^{3}}\!dk_{1}
\left(\,|k_{1}|^{-2}\,\sum_{\nu,j}\left(\;(1+|k_{1}|^{-3})|\hat{\varphi}_{j}^{\nu}(k_{1})|^{2}+\sum_{\lambda}|k_{1}|^{-1}\,|\,\partial_{\lambda}\hat{\varphi}_{j}^{\nu}  (k_{1})\,|^{2}\right)\,\right)^{p/2}\nonumber \\
&&<\infty,
\end{eqnarray}
where $C'$ is a constant independent of $m$.
Then we see that
\begin{eqnarray}
D_{k(i,\mu)}\Phi_{m}^{(n)}\in L^{2}(\mathbb{R}^{3+3n};\mathbb{C}^{2})\label{N3}
\end{eqnarray}
and
\begin{eqnarray}
\Vert D_{k(i,\mu)}\Phi_{m}^{(n)}\Vert_{p}<C'',\label{N4}
\end{eqnarray}
where $C''$ is a constant independent of sufficiently small $m$.
Next we consider the distributional derivative of $\Phi_{m}^{(n)}$ in $x$.
Since $\Phi_{m}\in D(|p|\otimes 1)$, the distributional derivative of $\Phi_{m}^{(n)}$ in $x$
is a function and for $n\geq 0$,
\begin{eqnarray}
\Vert\, D_{x,\mu}\, \Phi_{m}^{(n)} \,\Vert_{L^{2}(\mathbb{R}^{3+3n})}^{2}
\leq (\,(p_{\mu}\otimes 1)\,\Phi_{m},\,(p_{\mu}\otimes 1)\,\Phi_{m}\,)\leq C'''<\infty\label{N5}
\end{eqnarray}
holds, where $C'''$ is a constant independent of sufficiently small $m$. Therefore the lemma follows from (\ref{N3}), (\ref{N4}) and (\ref{N5}). \qed
\end{pf}
\subsection{Massless case}
{\it Proof of Theorem \ref{main 2}:}
By the Alaoglu theorem, there exists a sequence  $\{m(j)\}_{j=1}^{\infty}$ going to $0$
so that $\{\Phi_{m(j)}\}_{j=1}^{\infty}$ converges weakly to some vector $\Phi$ in $Q(\hat{q}^{V})$.
It is enough to show that $\{\Phi_{m(j)}\}_{j=1}^{\infty}$ strongly converges to $\Phi$
in $\mathcal{H}$. (See \cite[Proof of Theorem 2.1]{GLL}.) Let $\epsilon>0$ be an arbitrary number. 
By Lemma \ref{number b'dd}, we can see that
\begin{eqnarray}
\Vert\,\Phi_{m(j)}-\Phi\,\Vert^{2}\leq \sum_{n=0}^{M}\Vert\,\Phi_{m(j)}^{(n)}-\Phi^{(n)}\,\Vert^{2}+\frac{4}{M}\sup_{j}\Vert\,N^{1/2}\,\Phi_{m(j)}\,\Vert^{2}. \label{m.2}
\end{eqnarray}
Let $B_n\subset \mathbb{R}^{3n}$ $(n=1,2,...,M)$ be sufficiently large balls centered at the origin with radius $L$. 
Since $\Phi_m^{(n)}(x,k_1,...,k_n)$ is symmetric with respect to $k_1,...,k_n$, by Lemma $\ref{number b'dd}$ we have
\begin{align}\label{m.3}
&\Vert \Phi_m^{(n)} \Vert_{L^2(\mathbb{R}^{3+3n})}^2= \frac{1}{n} \int_{\mathbb{R}^3} \Vert (a\Phi_m)(k)^{(n)} \Vert^2 dk \nonumber \\
&\leq \int_{|k|<L} \Vert \Phi_m^{(n)}(\cdot,k,\cdot)\Vert_{L^2(\mathbb{R}^{3+3(n-1)})}^2 dk
+ C\alpha \,\sum_{\mu,j}\,\int_{|k|\geq L}\, \frac{|\hat{\varphi}_{j}^{\mu}(k)|^{2}\,\theta_{j}(k)^{2}}{\omega(k)\,\omega_{m}(k)^{2}}\,dk
 \nonumber\\
&\leq \int_{|k|<L} \Vert \Phi_m^{(n)}(\cdot,k,\cdot)\Vert_{L^2(\mathbb{R}^{3+3(n-1)})}^2 dk
 + \frac{\epsilon}{6nM}
 \leq  \Vert \Phi_m^{(n)} \Vert_{L^2(\mathbb{R}^3\times B_n)}^2+\frac{\epsilon}{6M}.
\end{align}
Furthermore by Theorem \ref{exp}, for all $R>0$,
\begin{align}
\Vert\,\Phi_{m(j)}^{(n)}-\Phi_{m(k)}^{(n)}\,\Vert^{2}
\leq\Vert(\Phi_{m(j)}^{(n)}-\Phi_{m(k)}^{(n)})\chi_{\{|x|<R\}}\Vert^{2} 
 +e^{-2c R}\sup_{j}\Vert e^{c |x|}\Phi_{m(j)}\Vert^{2}
\label{m.4}
\end{align}
holds.
Pick sufficiently large ball $\Omega$ centered at the origin with radius $R$ and let $M$ be a sufficiently large number. Then by (\ref{m.2}), (\ref{m.3}) and (\ref{m.4}), we have
\begin{align}
&\Vert\,\Phi_{m(j)}-\Phi_{m(k)}\,\Vert^{2} \nonumber \\
&\leq \sum_{n=0}^{M} \Vert\, \Phi^{(n)}_{m(j)} -\Phi^{(n)}_{m(k)}\, \Vert^{2}_{ L^{2}(\Omega_{n}) }
+\frac{2}{M}\sup_{j}\Vert\,N^{1/2}\,\Phi_{m(j)}\,\Vert^{2}+e^{-2c R}\,\sup_{j}\Vert\,e^{c |x|}\,\Phi_{m(j)}^{(n)}\,\Vert^{2} \nonumber \\
&<\sum_{n=0}^{M} \Vert\, \Phi^{(n)}_{m(j)} -\Phi^{(n)}_{m(k)}\, \Vert^{2}_{ L^{2}(\Omega_{n}) }+\frac{\epsilon}{2}. \label{m.5}
\end{align}
Since $\{\Phi_{m(j)}^{(n)}\}_{j=1}^{\infty}$ is bounded in $W^{1,p}(\Omega_{n})$ with $1\leq p<2$,
we see that $\Phi_{m(j)}^{(n)}$ weakly converges to $\Phi^{(n)}$ in $W^{1,p}(\Omega_{n})$.
Thus the Rellich-Kondrachov theorem \cite{Analysis} implies that $\Phi_{m(j)}^{(n)}$ strongly converges to $\Phi^{(n)}$.
Therefore for sufficiently large numbers $j$ and $k$, we have
\begin{eqnarray}
\Vert\,\Phi_{m(j)}-\Phi_{m(k)}\,\Vert^{2} <\epsilon.\label{m.6}
\end{eqnarray}
Since by (\ref{m.5}) and (\ref{m.6}), $\{\Phi_{m(j)} \}_{j=1}^{\infty}$ is a Cauchy sequence in $\mathcal{H}$ and
$\{\Phi_{m(j)}\}_{j=1}^{\infty}$ weakly converges to $\Phi$ in $\mathcal{H}$, we see that
\begin{eqnarray}
\lim_{j\to\infty}\Vert\,\Phi_{m(j)}-\Phi\,\Vert= 0.
\end{eqnarray}
Thus the theorem follows. \qed

\section{Appendix}
\subsection{Appendix A}
Lemmas \ref{mgll}, \ref{locally.3}, \ref{bbb} and Corollary \ref{k} will be proven in this subsection.
These proofs have been applied to standard the Pauli-Fierz Hamiltonian.

{\it Proof of Lemma \ref{mgll}:}
Let $\{ \Phi^{j}\}\subset Q(\hat{q}_{m}^{V})$ be a minimizing sequence of $\hat{q}_{m}^{V}$:
\begin{eqnarray}
\Vert \,\Phi^{j}\, \Vert =1,\quad \hat{q}_{m}^{V}(\,\Phi^{j},\,\Phi^{j}\,)\to E^{V}_{m} \text{ as }j\to \infty.
\end{eqnarray}
Using the Alaoglu theorem, we can suppose that
$\Phi^{j}\rightharpoonup \Phi_{m}$ weakly in $Q(\hat{q}_{m}^{V})$ by passing a subsequence.
Put $\tilde{\Psi}^{j}=\Phi_{m}-\Phi^{j}$. Then
\begin{eqnarray*}
0=\lim_{j\to \infty }\,\left(\hat{q}_{m}^{V}(\,\Phi^{j},\,\Phi^{j}\,)-E^{V}_{m}(\,\Phi^{j},\,\Phi^{j}\,) \right)
=\lim_{j\to\infty }\,\left(\hat{q}_{m}^{V}(\tilde{\Psi}^{j},\tilde{\Psi}^{j})-E^{V}_{m}(\tilde{\Psi}^{j},\tilde{\Psi}^{j})\right) \\
+\hat{q}_{m}^{V}(\,\Phi_{m},\,\Phi_{m}\,)-E^{V}_{m}(\,\Phi_{m},\,\Phi_{m}\,).
\end{eqnarray*}
This implies that
\begin{eqnarray}
\lim_{j\to\infty }\,(\,\hat{q}_{m}^{V}(\,\tilde{\Psi}^{j},\,\tilde{\Psi}^{j}\,)-E^{V}_{m}(\,\tilde{\Psi}^{j},\,\tilde{\Psi}^{j}\,)\,)=0 \label{bc}
\end{eqnarray}
and
\begin{eqnarray}
\hat{q}_{m}^{V}(\,\Phi_{m},\,\Phi_{m}\,)-E^{V}_{m}(\,\Phi_{m},\,\Phi_{m}\,)=0.
\end{eqnarray}
Thus it suffices to prove that $\Phi_{m}\neq 0$.
Suppose  that $\inf_{j} \Vert \tilde{\Psi}^{j} \Vert > 0$ and put $\Psi^{j}=\tilde{\Psi}^{j}/\Vert \tilde{\Psi}^{j}\Vert$.
Since for all $\eta\in\mathcal{H}$, $|(\eta,\Psi^{j})|\leq (1/\inf\Vert\tilde{\Psi}^{j}\Vert)\,|(\eta,\tilde{\Psi}^{j})|\rightarrow 0$, $\Psi^{j}\rightharpoonup 0$ weakly in $\mathcal{H}$.
Therefore by Lemma \ref{m>0} and (\ref{bc}), we see that
\begin{eqnarray*}
0<\liminf_{j\to\infty}\left(\hat{q}_{m}^{V}(\Psi^{j},\Psi^{j})-E^{V}_{m}\right)
\leq \frac{\dis\lim_{j\to\infty} (\hat{q}_{m}^{V}(\tilde{\Psi}^{j},\tilde{\Psi}^{j})-E^{V}_{m}(\tilde{\Psi}^{j},\tilde{\Psi}^{j}))}
{\dis\inf_{j}(\,\tilde{\Psi}^{j},\tilde{\Psi}^{j})}=0.
\end{eqnarray*}
This is a contradiction. Thus $\inf_{j}\Vert \tilde{\Psi}^{j} \Vert = 0$ and hence $\Vert\Phi_{m}\Vert=1$ follows.\qed

{\it Proof of Lemma \ref{locally.3}}:
It is  seen that on $\mathcal{S}$,
\begin{eqnarray}
(\,1_{\mathcal{H}}\otimes P_{0})\,)(\,1_{\mathcal{H}_{p}}\otimes U_{P}\,)=1_{\mathcal{H}_{p}}\otimes \Gamma(\,\hat{j}_{P,1}\,). \label{f}
\end{eqnarray}
Since operators of both sides of (\ref{f}) are bounded, (\ref{f}) can be extended on $\mathcal{H}$ and we find that
\begin{eqnarray}
\Vert \,(\, 1_{\mathcal{H}}\otimes P_{0})\,)\,(\,\phi_{R}\otimes U_{P}\,)\,\Psi^{j} \,\Vert = \Vert\, \phi_{R}\otimes\Gamma(\,\hat{j}_{P,1}\,)\,\Psi^{j}\, \Vert.
\end{eqnarray}
Let $\Pi_{n}$ be the projection from $\mathcal{F}$ to $\mathcal{F}_{\leq n} = \{\Psi \in \mathcal{F}\,|\, \Psi^{(k)}= 0 ,\; k\geq n\,\}$.
Note that $\{(1\otimes N^{1/2})\Psi^{j}\}_{j}$ is bounded since $\{\Vert\Psi^{j}\Vert_{+1}\}_{j=1}^{\infty}$ is bounded. Then we can see that
\begin{eqnarray}
\lefteqn{\Vert\, (\phi_{R} \otimes \Gamma(\,\hat{j}_{P,1}\,))\Psi^{j}\, \Vert^{2}}\nonumber \\
&&=\Vert\, (\phi_{R} \otimes \Gamma(\,\hat{j}_{P,1}\,))(1\otimes \Pi_{n})\Psi^{j}\, \Vert^{2}
+\Vert\,(\,\phi_{R}\otimes\Gamma(\hat{j}_{P,1})\,)\,(1\otimes (1-\Pi_{n}))\Psi^{j}\,\Vert^{2}\nonumber \\
&&\leq \Vert\, (\phi_{R} \otimes \Gamma(\,\hat{j}_{P,1}\,))(1\otimes \Pi_{n})\Psi^{j}\, \Vert^{2}
+n^{-1}\sup_{j}\Vert\,(1\otimes N^{1/2})\Psi^{j}\,\Vert^{2}.\label{aaaa}
\end{eqnarray}
The first term of the right hand side of (\ref{aaaa}) can be written as
\begin{eqnarray}
\lefteqn{\Vert\, (\phi_{R} \otimes \Gamma(\,\hat{j}_{P,1})\Pi_{n})\Psi^{j}\, \Vert^{2}} \nonumber \\
&&=\Big\Vert\left(\phi_{R}(\,1+p^{2}\,)^{-1/4}\;\otimes \; \Gamma(\,\hat{j}_{P,1}\,)\,(1+\hhfm  \,)^{-1/4}\Pi_{n}\right) \times \nonumber \\
&&\qquad\,(\,1+(p^{2}\otimes 1)\,)^{1/4}(\,1+(1\otimes \hhfm   )\,)^{1/4} (\,\hat{H}_{m}^{V}-E^{V}_{m}+1\,)^{-1/2}\times \nonumber \\
&&\qquad\,(\,\hat{H}_{m}^{V}-E^{V}_{m}+1\,)^{1/2}\, \Psi^{j} \,\Big\Vert^{2}.
\end{eqnarray}
Note that since $\{\Vert \Psi^{j}\Vert_{+1}\}_{j=1}^{\infty}$ is a bounded sequence and $\Psi^{j}\rightharpoonup 0$ \,weakly in $\mathcal{H}$,
we find that $\Psi^{j} \rightharpoonup 0 $ weakly in $Q(\hat{q}_{m}^{V})$.
Since $$\phi_{R}(1+p^{2})^{-1/4}\; \otimes \; \Gamma(\,\hat{j}_{P,1}\,)(\,1+\hhfm  \,)^{-1/4}\Pi_{n}$$ is compact
and $$ (1+p^{2})^{1/4}\otimes(1+\hhfm  )^{1/4}(\,\hat{H}_{m}^{V}-E^{V}_{m}+1\,)^{-1/2}$$ is bounded, we can see that
\begin{eqnarray}
\lefteqn{\Big\Vert \left(\phi_{R}(\,1+p^{2}\,)^{-1/4}\;\otimes \; \Gamma(\,\hat{j}_{P,1}\,)\,(1+\hhfm  \,)^{-1/4}\Pi_{n}\right)\times}  \nonumber \\
&&(1+p^{2})^{1/4}\otimes(1+ \hhfm   )^{1/4}\,) (\,\hat{H}_{m}^{V}-E^{V}_{m}+1\,)^{-1/2}\times \nonumber \\
&&(\,\hat{H}_{m}^{V}-E^{V}_{m}+1\,)^{1/2}\, \Psi^{j} \,\Big\Vert^{2}\rightarrow 0 \; \text{ as }\;j\to\infty.
\end{eqnarray}
Therefore letting $j\to \infty $ in (\ref{aaaa}), we have
\begin{eqnarray}
\lim_{j\to\infty} \Vert\, (1 \otimes \Gamma(\,\hat{j}_{P,1}\,))\Psi^{j}_{R}\, \Vert^{2} \leq n^{-1}\sup_{j}\,\Vert\,(1\otimes N^{1/2})\Psi^{j}\Vert.
\end{eqnarray}
Letting $n\to\infty$, we can see that the lemma follows.
\qed

{\it Proof of Lemma \ref{bbb}}:
Note that as $m$ is decreasing, $\hat{q}_{m}^{V}$ is monotonously nonincreasing in the sense of form.
Then there exists $E^{*}$ so that $E^{V}_{m}\downarrow E^{*}$ as $m \downarrow 0$ and $E^{*}\geq E^{V}_{0}$.
Let $\Phi \in Q(\hat{q}^{V})$ be a normalized vector satisfying
\begin{eqnarray}
\hat{q}^{V}(\Phi,\Phi)<E^{V}_{0}+\epsilon .
\end{eqnarray}
Since \[ \lim_{n\to \infty} \hat{q}^{V}\left(\,\frac{(1\otimes \Pi_{n})\Phi}{\Vert \,(1\otimes \Pi_{n})\Phi\, \Vert},\,\frac{(1\otimes \Pi_{n})\Phi}{\Vert \,(1\otimes \Pi_{n})\Phi\,\Vert}\right)=\hat{q}^{V}(\Phi,\,\Phi) ,\]
we can take $\Phi\in Q(\hat{q}^{V})\cap D(1\otimes N)$ such that
$$E^{V}_{m}\leq \hat{q}_{m}^{V}(\Phi,\Phi)\leq \hat{q}^{V}(\Phi,\Phi )+m( \Phi,\, (1\otimes N) \Phi )<E^{V}_{0}+\epsilon+m( \Phi,\,(1\otimes N) \Phi ).$$
Letting $m\downarrow 0$, we have $E^{*}= E^{V}_{0}$ and the lemma follows.\qed

{\it Proof of Corollarly \ref{k}}: 
By binding condition for $m=0$, there exists $\epsilon >0$ satisfying that
\begin{eqnarray}
E^{V}_{0}<\lim_{R\to \infty}E^{V}_{R,0}-2\epsilon. \label{k.1}
\end{eqnarray}
By Lemma \ref{bbb} and (\ref{k.1}), we have for sufficiently small $m$,
\begin{eqnarray}
E^{V}_{m}<E^{V}_{0}+\epsilon<\lim_{R\to \infty}E^{V}_{R,0}-\epsilon \leq \lim_{R\to \infty}E^{V}_{R,m}-\epsilon.
\end{eqnarray}
Thus the corollary follows. \qed
\subsection{Appendix B}
\begin{prop}\label{app.a}
Suppose that $j\in C^{2}(\mathbb{R}^{3})$ and $\nabla j$ is bounded. Set $j_{P}=j(\cdot/P)$ for $P>1$
and $\hat{j}_{P}=j_{P}(-i\nabla )$.
Let $m>0$.
Then $[\omega_{m},\hat{j}_{P}]$ is a bounded operator and
\begin{eqnarray}
\Vert \,[\omega_{m},\hat{j}_{P}]\, \Vert < \frac{C}{P},
\end{eqnarray}
where $C$ is a constant depending on $m$.
\end{prop}
\begin{pf}
$[\omega_{m},\hat{j}_{P}]$ can be written as
\begin{eqnarray}
[\omega_{m},\hat{j}_{P}]=\frac{1}{\omega_{m}} [\omega_{m}^{2},\hat{j}_{P}]+\left[ \frac{1}{\omega_{m}} , \hat{j}_{P}\right] \omega_{m}^{2}. \label{ww}
\end{eqnarray}
The first term of (\ref{ww}) can be estimated as
\begin{eqnarray}
\left\Vert \frac{1}{\omega_{m}} [\omega_{m}^{2},\hat{j}_{P}]\right\Vert \leq \frac{\mathrm{const.}}{P}\left(\frac{1}{m}+1\right).
\end{eqnarray}
Let us consider the second term of $(\ref{ww})$.
$\omega_{m}^{-1}$ can be written as
\begin{eqnarray}
\frac{1}{\omega_{m}}=\frac{2}{\pi}\int_{0}^{\infty} \frac{1}{t^{2}+\omega_{m}^{2}}dt,
\end{eqnarray}
where the integral is $B(L^{2}(\mathbb{R}^{3}))$-valued Riemannian integral.
Then for $\Psi,\Phi\in C_{c}^{\infty}(\mathbb{R}^{3})$,
\begin{eqnarray}
{\left( \Psi,\,\left[ \frac{1}{\omega_{m}},\, \hat{j}_{P} \right]\omega_{m}^{2}\Phi \right)}
&=&\frac{2}{\pi}\int_{0}^{\infty} \left( \Psi,\,\left[ \frac{1}{t^{2}+\omega_{m}^{2}},\,\hat{j}_{P} \right]\omega_{m}^{2}\Phi\right) dt\nonumber\\
&=&\frac{2}{\pi}\int_{0}^{\infty} \left( \Psi,\,\frac{k}{t^{2}+\omega_{m}^{2}}\cdot (-i\nabla \hat{j}_{P}) \,\frac{\omega_{m}^{2}}{t^{2}+\omega_{m}^{2}}\Phi\right) dt \nonumber \\
&&\qquad+\frac{2}{\pi}\int_{0}^{\infty} \left( \Psi,\,\frac{1}{t^{2}+\omega_{m}^{2}}      (-i\nabla \hat{j}_{P}) \cdot\frac{k\omega_{m}^{2}}{t^{2}+\omega_{m}^{2}}\Phi\right) dt\nonumber\\
&=&\frac{4}{\pi}\int_{0}^{\infty} \left( \Psi,\frac{k}{t^{2}+\omega_{m}^{2}}\cdot (-i\nabla \hat{j}_{P}) \,\frac{\omega_{m}^{2}}{t^{2}+\omega_{m}^{2}}\Phi\right) dt\nonumber \\
&&\qquad  +\frac{2}{\pi}\int_{0}^{\infty} \left( \frac{1}{t^{2}+\omega_{m}^{2}}\Psi,(  \Delta \hat{j}_{P}) \frac{\omega_{m}^{2}}{t^{2}+\omega_{m}^{2}}\Phi\right) dt.\label{www}
\end{eqnarray}
Here $\nabla \hat{j}_{P}= (\nabla j_{P})(-i\nabla)$, $\Delta\hat{j}_{P}=(\Delta j_{P})(-i\nabla)$ and $k=(k_{1},k_{2},k_{3})$ is a multiplication operator.
The second term of the right hand side of (\ref{www}) can be estimated as
\begin{eqnarray}
\left|\,\frac{2}{\pi}\int_{0}^{\infty} \left( \frac{1}{t^{2}+\omega_{m}^{2}}\Psi,(  \Delta \hat{j}_{P}) \frac{\omega_{m}^{2}}{t^{2}+\omega_{m}^{2}}\Phi\right) dt\,\right|
\leq \frac{\mathrm{const.}}{P^{2}m},
\end{eqnarray}
and the first term is
\begin{eqnarray}
\lefteqn{\frac{4}{\pi}\int_{0}^{\infty} \left( \Psi,\frac{k}{t^{2}+\omega_{m}^{2}}\cdot (-i\nabla \hat{j}_{P}) \left( 1-\frac{t^{2}}{t^{2}+\omega_{m}^{2}}\right) \Phi\right) dt\nonumber}\\
&=&\frac{4}{\pi}\int_{0}^{\infty} \left( \left( 1- \frac{ t^{2} }{ t^{2} + \omega_{m}^{2} } \right) \frac{ k }{ t^{2} + \omega_{m}^{2} }\Psi,  (-i\nabla \hat{j}_{P}) \Phi \right) dt \nonumber\\
&&\quad +\frac{4}{\pi}\int_{0}^{\infty} \left( \Psi,\frac{t^{2}}{t^{2} + \omega_{m}^{2}}\frac{k}{t^{2}+\omega_{m}^{2}}\cdot [(-i\nabla \hat{j}_{P}),\,k^{2}]\frac{1}{t^{2}+\omega_{m}^{2}} \Phi \right) dt \nonumber \\
&=&\frac{4}{\pi}\int_{0}^{\infty} \left( \left( 1- \frac{ t^{2} }{ t^{2} + \omega_{m}^{2} } \right) \frac{ k }{ t^{2} + \omega_{m}^{2} }\Psi,  (-i\nabla \hat{j}_{P})  \Phi \right) dt \nonumber \\
&&\quad+\sum_{\mu=1}^{3}\frac{4}{\pi}\int_{0}^{\infty} \left( \frac{1}{t^{2} + \omega_{m}^{2}}\frac{k_{\mu}k}{t^{2}+\omega_{m}^{2}}\Psi, (i\partial_{\mu}\nabla \hat{j}_{P})\frac{t^{2}}{t^{2}+\omega_{m}^{2}} \Phi \right) dt\nonumber\\
&&\quad+\sum_{\mu=1}^{3}\frac{4}{\pi}\int_{0}^{\infty} \left( \frac{t^{2}}{t^{2} + \omega_{m}^{2}}\frac{k}{t^{2}+\omega_{m}^{2}}\Psi, (i\partial_{\mu}\nabla \hat{j}_{P})\frac{k_{\mu}}{t^{2}+\omega_{m}^{2}} \Phi \right) dt.\quad \quad
\end{eqnarray}
The first term can be estimated as
\begin{eqnarray}
\lefteqn{\left| \frac{4}{\pi}\int_{0}^{\infty} \left( \left( 1- \frac{ t^{2} }{ t^{2} + \omega_{m}^{2} } \right) \frac{ k }{ t^{2} + \omega_{m}^{2} }\Psi,  (-i\nabla \hat{j}_{P}) \Phi \right) dt\right| }\nonumber\\
&\leq& \frac{4}{\pi} \int_{\mathbb{R}^{3}}\int_{0}^{\infty} \left| \left( 1- \frac{ t^{2} }{ t^{2} + \omega_{m}^{2} } \right) \frac{ k }{ t^{2} + \omega_{m}^{2} } \Psi(k) \right| \, | (-i\nabla \hat{j}_{P}) \Phi (k)| dtdk\nonumber \\
&\leq& 2\int_{\mathbb{R}^{3}}|\Psi(k) | \, | (-i\nabla \hat{j}_{P})  \Phi (k)|dk\leq \frac{\mathrm{const.}}{P}\Vert\Psi\Vert\, \Vert\Phi\Vert.
\end{eqnarray}
Other terms can be estimated in a similar way and we obtain that
\begin{eqnarray}
\left|\left( \Psi,\left[ \frac{1}{\omega_{m}}, \hat{j}_{P} \right]\omega_{m}^{2}\Phi \right)\right|
\leq\frac{C}{P}\Vert \Psi \Vert \,\Vert \Phi \Vert.
\end{eqnarray}
Here $C$ depends on $m$.\qed
\end{pf}
\subsection{Appendix C}
\begin{lem}
Let $\theta=(\theta_{1},\theta_{2})\in L^{\infty}(\mathbb{R}^{3};\mathbb{C}^{2})$ and $\{g_{i}\}_{i=1}^{\infty}$ a complete orthonormal system of $L^{2}(\mathbb{R}^{3};\mathbb{C}^{2})$.
Suppose that $g_{i}\in D(\omega_{m}^{1/2})$. Then
\begin{eqnarray}
\sum_{i}\left(a(\omega_{m}^{-1/2}\theta g_{i})\Phi_{m},\,a(\omega_{m}^{1/2}\theta g_{i})\Phi_{m}\right)
=\sum_{j=1,2}\int_{\mathbb{R}^{3}}\theta_{j}(k)^{2}\Vert (a_{j}\Phi_{m})(k)\Vert^{2}dk \label{6.2}
\end{eqnarray}
holds.
\end{lem}
\begin{pf}
We set that
\begin{eqnarray}
\lefteqn{\sum_{i}\,\left(\,a(\omega_{m}^{-1/2}\theta\, g_{i})\,\Phi_{m},\,a(\omega_{m}^{1/2}\theta\,g_{i})\,\Phi_{m} \right)}\nonumber \\
&&=\sum_{i}\sum_{n=0}^{\infty}\int dk_{1}\ldots dk_{n}(n+1)
\overline{\int \omega_{m}^{-1/2}(k)\,\theta(k )g_{i}(k )\Phi_{m} (k,k_{1},\ldots,k_{n})dk} \nonumber \\
&&\hspace{4.5cm} \int  \omega_{m}^{ 1/2}(k')\theta(k')g_{i}(k')\Phi_{m} (k',k_{1},\ldots,k_{n})dk'\nonumber \\
&&=\sum_{i}\sum_{n=0}^{\infty}\,\int \,dk_{1}\ldots dk_{n} \,(n+1)\nonumber \\
&&\hspace{9mm}\left(\,\overline{\omega_{m}^{1/2}\theta\Phi_{m}^{(n+1)}(\cdot,k_{1},\ldots,k_{n})}  ,\, g_{i}\right)
\left(g_{i},\,\overline{\omega_{m}^{-1/2}\theta\Phi_{m}^{(n+1)}(\cdot,k_{1},\ldots,k_{n})}\right). \nonumber
\end{eqnarray}
Note that by Bessel's inequality, we can see that
\begin{eqnarray}
\lefteqn{\left|\sum_{i=1}^{N}(n+1)\left(\overline{\omega_{m}^{1/2}\theta \Phi_{m}^{(n+1)}(\cdot,k_{1},\ldots,k_{n})},g_{i}\right)
\left(g_{i},\overline{\omega_{m}^{-1/2}\theta\Phi_{m}^{(n+1)}(\cdot,k_{1},\ldots,k_{n})}\right) \right|}\nonumber\\
&&\leq \frac{1}{2}\,\Big(\,\left\Vert \, (n+1)^{1/2}\omega_{m}^{1/2}\theta\,\Phi_{m}^{(n+1)}(\cdot,k_{1},\ldots,k_{n})\,\right\Vert^{2}\nonumber \\
&&\hspace{2cm} +\left\Vert \,(n+1)^{1/2}\,\omega_{m}^{-1/2}\,\theta\,\Phi_{m}^{(n+1)}(\cdot,k_{1},\ldots,k_{n}) \,\right\Vert^{2} \, \Big).\hspace{4cm}\nonumber
\end{eqnarray}
Since  $\Phi_{m}\in D(d\Gamma(\omega_{m})^{1/2})$, the right hand side above is summable for $n,k_{1}\ldots k_{n}$.
Thus, by the Lebesgue dominated convergence theorem and Parseval's equality, we obtain $(\ref{6.2})$.\qed
\end{pf}
{\footnotesize
\section*{Acknowledgments}
I thank Prof. F. Hiroshima for his helpful advice.

}\end{document}